\documentclass[10pt,prl,aps,twocolumn]{revtex4-2}
\usepackage{color,ifthen,amsthm,amsmath,amsxtra,amsfonts,dsfont,graphicx,bm,tikz,scalerel,wasysym,bbm,graphicx,amsthm,braket,physics,empheq, comment}
\usepackage[colorlinks=true, linkcolor=blue, citecolor=blue, urlcolor=blue, bookmarks]{hyperref}
\usepackage[inline]{enumitem}
\usepackage{CircuitTikz}

\newcommand{\be}{\begin{equation}}
\newcommand{\ee}{\end{equation}}
\newcommand{\bea}{\begin{eqnarray}}
\newcommand{\eea}{\end{eqnarray}}
\newcommand{\bse}{\begin{subequations}}
\newcommand{\ese}{\end{subequations}}
\def\Id{{\openone}}

\newcommand{\SoPA}{School of Physics and Astronomy, University of Nottingham, Nottingham, NG7 2RD, UK}

\newcommand{\CQNE}{Centre for the Mathematics and Theoretical Physics of Quantum Non-Equilibrium Systems, University of Nottingham, Nottingham, NG7 2RD, UK}

\begin{document}

\title{Exact quench dynamics of the Floquet quantum East model at the deterministic point}

\author{Bruno Bertini}
\affiliation{\SoPA} \affiliation{\CQNE}

\author{Cecilia De Fazio}
\affiliation{\SoPA} \affiliation{\CQNE}

\author{Juan P.\ Garrahan}
\affiliation{\SoPA} \affiliation{\CQNE}

\author{Katja Klobas}
\affiliation{\SoPA} \affiliation{\CQNE}


\begin{abstract}
  We study the non-equilibrium dynamics of the Floquet quantum East model (a Trotterized version of the kinetically constrained quantum East spin chain) at its ``deterministic point'', where evolution is defined in terms of CNOT permutation gates. We solve exactly the thermalization dynamics for a broad class of initial product states by means of ``space evolution''. We prove: (i) the entanglement of a block of spins grows at most at one-half the maximal speed allowed by locality (i.e., half the speed of dual-unitary circuits); (ii) if the block of spins is initially prepared in a classical configuration, speed of entanglement is a quarter of the maximum; (iii) thermalization to the infinite temperature state is reached exactly in 
  a { time that scales with the size of the block.}
\end{abstract}

\maketitle

\noindent
{\textit{Introduction.---}} Systems with constrained dynamics are of interest in many areas of non-equilibrium physics. Kinetically constrained models (KCMs) \cite{fredrickson1984kinetic,jackle1991a-hierarchically,ritort2003glassy} provide a framework for explaining \cite{garrahan2002geometrical,chandler2010dynamics,speck2019dynamic} the emergence of slow and heterogeneous dynamics in glasses \cite{berthier2011theoretical,biroli2013perspective,garrahan2018aspects,hasyim2023emergent}, and their study has spurred the development of dynamical large deviation and trajectory ensemble methods \cite{touchette2009the-large,touchette2018introduction,jack2020ergodicity}. Quantum constrained dynamics emerges naturally in systems such as Rydberg atoms under blockade conditions \cite{lesanovsky2011many-body, bernien2017probing, browaeys2020many-body, bluvstein2021controlling}, leading to questions about slow thermalisation and non-ergodicity in the absence of disorder \cite{turner2018weak, ho2019periodic, emergent2019choi, sala2020ergodicity, zadnik2021the, zadnik2021the2, pozsgay2021integrable, singh2021subdiffusion, serbyn2021quantum, moudgalya2022quantum, singh2023fredkin, gopalakrishnan2023distinct, ljubotina2023superdiffusive, brighi2023hilbert}. 

The simplest setting for implementing kinetic constraints is in lattice systems with discrete dynamics, such as cellular automata \cite{wolfram1983statistical,bobenko1993two} or quantum circuits~\cite{osborne2006efficient}. For such setups it has been possible to obtain many exact results that underpin our understanding of quantum dynamics, 
  { including on operator dynamics, information spreading, and thermalisation (see e.g.\ Refs.~\cite{nahum2017quantum, vonKeyserlingk2018operator,  chan2018solution, khemani2018operator, rakovszky2018diffusive, zhou2020entanglement,reid2021entanglement, wang2019barrier, friedman2019spectral, bertini2018exact,chan2018spectral, flack2020statistics, bertini2021random, fritzsch2021eigenstate, kos2021thermalization, bertini2022exact, garratt2021manybody, garratt2021local, piroli2020exact, claeys2021ergodic, suzuki2022computational,klobas2021exact, klobas2021exact2, klobas2021entanglement,kos2021thermalization,nahum2017quantum,nahum2018operator,chan2018solution,vonKeyserlingk2018operator,bertini2019entanglement,bertini2019exact,friedman2019spectral,li2019measurement,skinner2019measurement,rakovszky2019sub,zabalo2020critical,claeys2020maximum}).
  Quantum circuits are also vital for experimental simulation of quantum systems and quantum computation, having been used to demonstrate quantum advantage, perform randomised benchmarking, and to study non-equilibrium Floquet dynamics~\cite{brydges_probing_2019,elben_many-body_2019,elben_renyi_2018,pichler_measurement_2016,vermersch_probing_2019,vermersch_unitary_2018, aaronson2018shadow,huang_predicting_2020,ohliger_efficient_2013, keenan2022evidence, morvan2022formation}. 
  Here we consider this setting to characterise the dynamical effects of kinetic constraints by studying} 
a circuit version of the quantum East model \cite{horssen2015dynamics,pancotti2020quantum,bertini2024localized},
itself a quantum generalisation of the classical East  model \cite{jackle1991a-hierarchically}. Using methods similar to those employed for dual-unitary circuits~\cite{bertini2019exact,bertini2019exact,piroli2020exact}, we solve exactly the thermalization dynamics. 

\begin{figure}
  \includegraphics[width=\columnwidth]{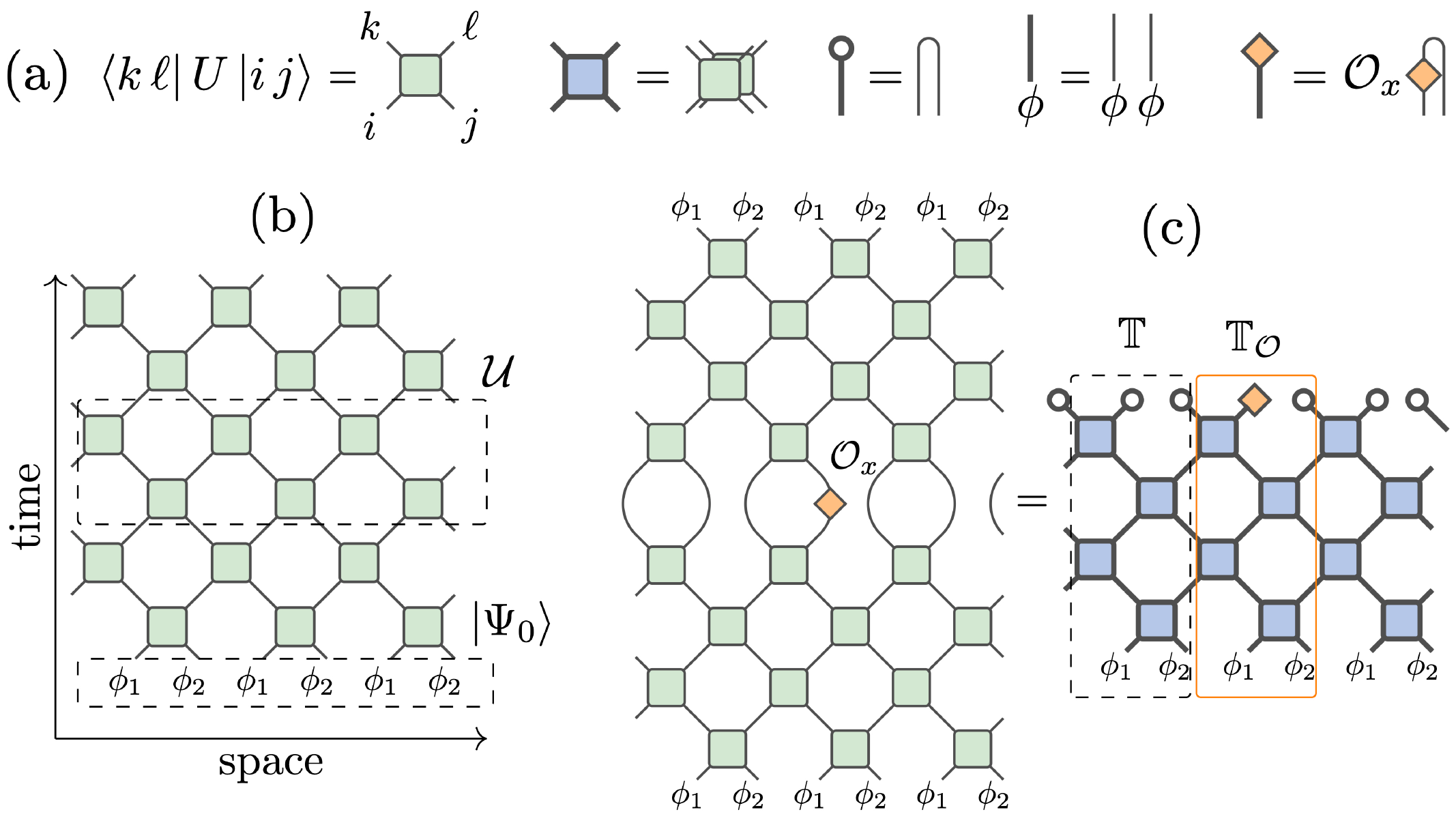}
  \caption{\label{fig:circuit}
    {Deterministic Floquet quantum East model.}  
    (a) Diagrammatic representation of the gate Eq.~\eqref{eq:localgate}. 
    Thick lines correspond to the folded representation of the forward and backward branches. 
    (b) 
    Time evolution of the state as a quantum circuit, 
    $\ket{\Psi(t)} = \mathcal{U}^t \, \ket{\Psi_0}$. The dashed boxes indicate the initial state $\ket{\Psi_0}$
    and the evolution operator ${\mathcal U}$ for one time step. 
    (c) One point function of a local operator (l.h.s.) and it folded representation (r.h.s.). The dashed and full boxes outline the space transfer matrices.
  }
\end{figure}

\noindent
{\textit{Model setting.---} }
More specifically, we consider the non-equilibrium dynamics of the Floquet Quantum East model~\cite{bertini2024localized} at its deterministic point, which we refer to as ``deterministic Floquet quantum East'' (DFQE) model. This system can be thought of as a brickwork quantum circuit, see Fig.~\ref{fig:circuit}, acting on a chain of $2L$ qubits and with local gate given by 
\begin{equation} \label{eq:localgate}
U = \Id \otimes \bar P + X \otimes P, 
\end{equation}
where $P= \Id-\bar P = (\Id+Z)/2$ is the projector to the up state $\ket{1}$ of the qubit (the down state is denoted by $\ket{0}$) and $\{X,Y,Z\}$ are Pauli matrices~\footnote{This gate can be recovered from the gate in Eq. (5) of Ref.~\cite{bertini2024localized} by performing a space inversion and setting $a=1$ and $\tau=\pi/2$}. The quantum circuit with local gate \eqref{eq:localgate} was first studied in Ref.~\cite{gopalakrishnan2018facilitated} { (see also Ref.~\cite{berenstein2021exotic})} and is the quantum counterpart of the classical Floquet East model of Ref.~\cite{klobas2023exact}.
The gate~\eqref{eq:localgate} deterministically implements the constraint that defines both the classical~\cite{jackle1991a-hierarchically,ritort2003glassy} and quantum~\cite{horssen2015dynamics,pancotti2020quantum,bertini2024localized} East models, where a site can flip only if its right neighbour is in the up state. The local gate \eqref{eq:localgate} is part of the so called second hierarchy of generalised dual-unitary circuits {(DU2)} introduced in Ref.~\cite{yu2024hierarchical}. In the jargon of quantum circuits, Eq.~\eqref{eq:localgate} is a CNOT (controlled NOT) gate~\cite{nielsen2010quantum}, { and as such it is a Clifford gate~\cite{gottesman1998heisenberg,gutschow2010time}. This implies that there exists a class of initial \emph{stabilizer states} whose dynamics can be efficiently simulated classically. Our discussion, however, is not restricted to this class.}

Following a standard quantum quench protocol~\cite{calabrese2006time}, the system is prepared in an initial state $\ket{\Psi_0}$, which we take to be a product state in space, and then let to evolve unitarily as in Fig.~\ref{fig:circuit}b. We characterise the ensuing dynamics using the so called {\em space evolution} approach (also known as folding algorithm)~\cite{banuls2009matrix} (see also Refs.~\cite{muellerhermes2012tensor, hastings2015connecting, sonner2022characterizing, frias2022light, lerose2023overcoming}). This can be used to characterise the evolution of general local observables~\cite{piroli2020exact, lerose2021influence, klobas2021exact, klobas2021exact2}, quantum information~\cite{bertini2019entanglement, ippoliti2021postselectionfree, ippoliti2021fractal, klobas2021entanglement, bertini2022entanglement, bertini2022growth, bertini2023nonequilibrium, bertini2023dynamics}, and even spectral properties~\cite{bertini2018exact, bertini2021random, flack2020statistics, garratt2021local, fritzsch2021eigenstate, garratt2021manybody}, but is most easily explained by considering the one-point function of an operator, $\mathcal O_x$, acting on a single qubit. We represent this via a tensor-network diagram and fold on the portions of the network representing forward and backward evolution, see Fig.~\ref{fig:circuit}c, and  contract the network horizontally (in space rather than time). Namely, we write the one-point function using the space transfer matrices $\mathbb T$ and $\mathbb T_{\mathcal O}$,  defined in Fig.~\ref{fig:circuit}c, as  
\be
\expval{\mathcal O_x}{\Psi(t)} = \tr[\mathbb T_{\mathcal O} \mathbb T^{L-1}]\,.
\label{eq:onepoint}
\ee
Note that the transfer matrices appearing in this expressions act on the vertical folded lattice, i.e., they act on the Hilbert space $\mathcal H_t = \left.\mathbb{C}^4\mkern-4mu\right.^{\otimes 2t}$, and for simplicity we assumed the initial state to be two-site shift invariant. This latter assumption is not necessary for our analysis and will be explicitly lifted in the second part of this work. 

Because of unitarity and locality of the interactions the transfer matrix $\mathbb T$ has a very simple spectrum: 
its only eigenvalues are $1$ and $0$~\cite{klobas2021exact2, bertini2022entanglement}. The transfer matrix itself is \emph{not} rank one, since the eigenvalue zero has generically a non-trivial Jordan structure. However, the size of its Jordan blocks are bounded by $2t$, which implies that $\mathbb{T}^{2t}$ \emph{is} rank one.
It can be written as $\mathbb T^{2t} = \ketbra{R}{L}$, where $\ket{R}$ and $\ket{L}$ are the right and left fixed points of $\mathbb T$~\footnote{We choose the normalisation such that $\braket{R}{L}=1$}. In other words, for $L\geq 2t$ the one-point function of interest is fully specified the fixed points. In particular we have 
\be
\lim_{L\to \infty}\expval{\mathcal O_x}{\Psi(t)} = \mel{L}{\mathbb T_{\mathcal O}}{R}\,.  
\label{eq:onepointTL}
\ee
This expression suggests an interesting physical interpretation of the fixed points: they are the mathematical objects encoding the influence of the rest of the system on the subsystem where $\mathcal O$ acts. For this reason they are also referred-to as \emph{influence matrices}~\cite{lerose2021influence}.

Equation~\eqref{eq:onepointTL} might seem to be a drastic simplification of Eq.~\eqref{eq:onepoint}, as it replaces a complicated matrix product with a matrix element between two fixed points. This form offers practical advantage only when the influence matrices can be computed efficiently, e.g., when they can be represented by matrix product states with low bond dimension. This is not possible in general: for generic systems and initial states, influence matrices have volume law entanglement in time \cite{foligno2023temporal}. Some systems, however, avoid this general rule. These includes a class of chaotic dual-unitary circuits~\cite{bertini2019exact} evolving from a family of compatible initial states~\cite{piroli2020exact}, and evolution from compatible states in the Rule 54 quantum cellular automaton~\cite{klobas2021exact, klobas2021exact2}. In fact, Ref.~\cite{giudice2022temporal} argued that in the presence of integrability every low entanglement initial state should generate low entangled influence matrices. Here we show that also the non-integrable DFQE admits \emph{solvable} initial states generating analytically tractable influence matrices. We find \emph{three} distinct families of initial states with influence matrices in dimer-product form, i.e., entangling together only pairs of sites along time. We then use this result to study the exact quench dynamics of a block of spins when the rest of the system is prepared in a solvable state. We show that, regardless of the initial state, the block relaxes to the infinite temperature state in a finite number of time steps. Moreover, we provide an exact description of the full entanglement dynamics if the block is initially prepared in a solvable state.

\begin{figure}
  \includegraphics[width=0.9\columnwidth]{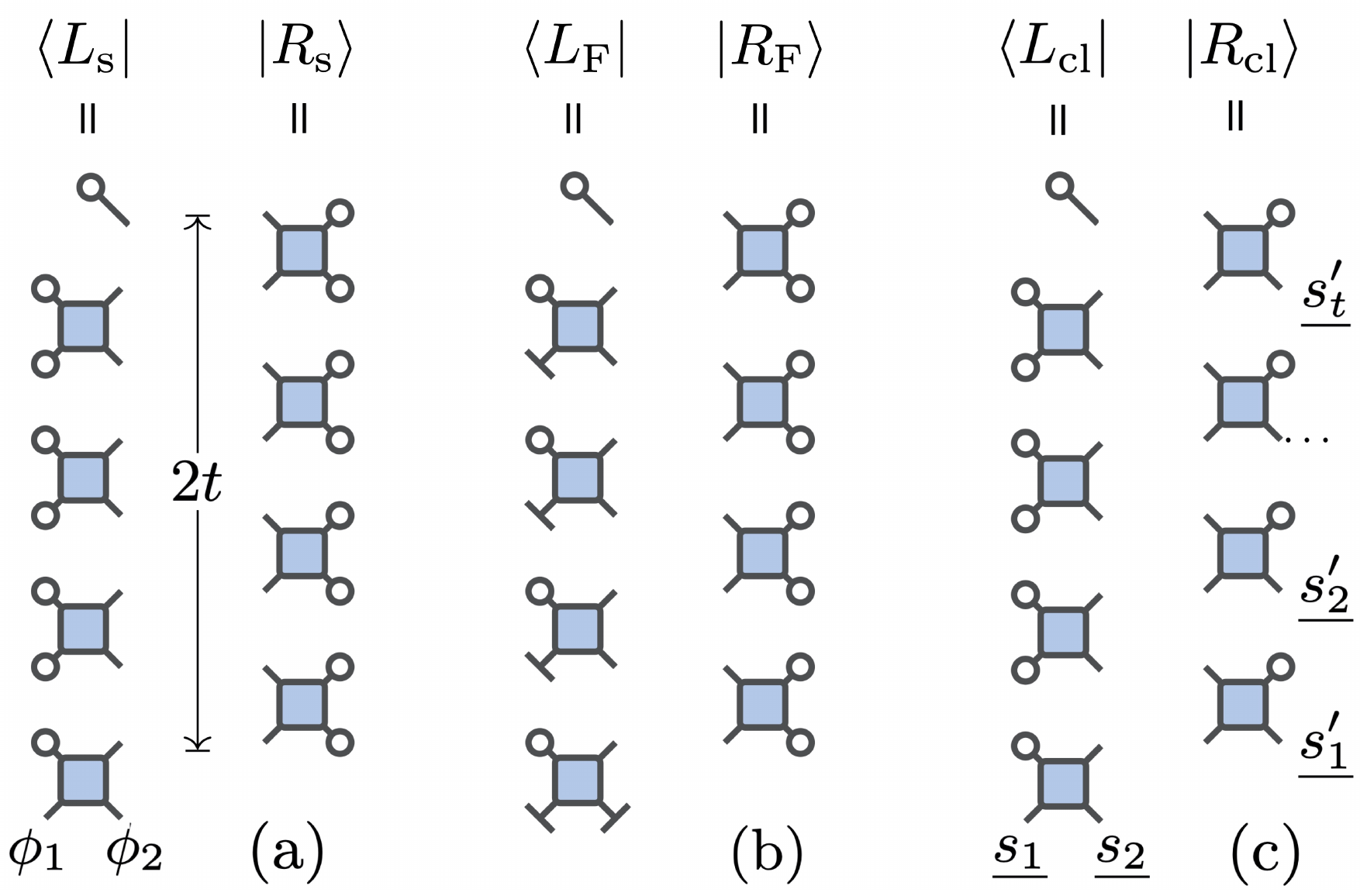}
    \caption{
      { Fixed points of space transfer matrices.}
      (a) Initial states $|\phi_{1,2} \rangle$ fulfilling Eq.~\eqref{eq:solvablestates}. 
      (b) Flat initial states, $\ket{\phi_{1,2}}=\ket{-}$. 
      (c) Classical initial states, in the notation of Eq.~\eqref{eq:sbar}, where ${s_1^\prime}\equiv{s_1 + s_2} {\pmod{2}}\,,\; {s_2^\prime}={s_2}$, and ${s_{t'}^\prime}\equiv{s^\prime_{t'-1} + s^\prime_{t'-2}} \pmod{2}$ for $2 < t' \leq t$ (see \cite{Note10} for details).
      }
    \label{fig:fixedpoints}
\end{figure}

\noindent
{\textit{Exact Fixed Points.---}} 
We begin by observing that $U$ in Eq.~\eqref{eq:localgate}, see Fig.~\ref{fig:circuit}, obeys the local relations
\be
  \begin{tikzpicture}[baseline={([yshift=-0.6ex]current bounding box.center)},scale=0.5]
    \prop{0}{0}{FcolU}
    \prop{1}{1}{FcolU}
    \fME{-0.5}{0.5}
    \fME{0.5}{1.5}
    \fME{-0.5}{-0.5}
  \end{tikzpicture}=
  \begin{tikzpicture}[baseline={([yshift=-0.6ex]current bounding box.center)},scale=0.5]
    \prop{1}{1}{FcolU}
    \gridLine{0.5}{-0.25}{1.5}{-0.25}
    \fME{0.5}{0.5}
    \fME{0.5}{1.5}
    \fME{0.5}{-0.25}
  \end{tikzpicture},
  \qquad 
    \begin{tikzpicture}[baseline={([yshift=-0.6ex]current bounding box.center)},scale=0.5]
    \prop{0}{0}{FcolU}
    \prop{-1}{1}{FcolU}
    \fME{0.5}{0.5}
    \fME{-0.5}{1.5}
    \fME{0.5}{-0.5}
  \end{tikzpicture}=
  \begin{tikzpicture}[baseline={([yshift=-0.6ex]current bounding box.center)},scale=0.5]
    \prop{1}{1}{FcolU}
    \gridLine{0.5}{-0.25}{1.5}{-0.25}
    \fME{1.5}{0.5}
    \fME{1.5}{1.5}
    \fME{1.5}{-0.25}
  \end{tikzpicture},
  \label{eq:LR}
\ee
which define {DU2} circuits \cite{yu2024hierarchical}. Relations \eqref{eq:LR} imply \cite{yu2024hierarchical} that if the initial states of two neighbouring sites fulfil 
\begin{equation}
  \label{eq:solvablestates}
  \begin{tikzpicture}[baseline={([yshift=-0.6ex]current bounding box.center)},scale=0.5]
    \prop{0}{0}{FcolU}
    \prop{1}{1}{FcolU}
    \fME{-0.5}{0.5}
    \fME{0.5}{1.5}
    \node at (0.5,-.85) {\scalebox{.8}{$\phi_2$}};
    \node at (-0.5,-.85) {\scalebox{.8}{$\phi_1$}};
  \end{tikzpicture}= \frac{1}{2}
  \begin{tikzpicture}[baseline={([yshift=-0.6ex]current bounding box.center)},scale=0.5]
    \prop{1}{1}{FcolU}
    \fME{0.5}{0.5}
    \fME{0.5}{1.5}
  \end{tikzpicture},
  \qquad 
    \begin{tikzpicture}[baseline={([yshift=-0.6ex]current bounding box.center)},scale=0.5]
    \prop{0}{0}{FcolU}
    \prop{-1}{1}{FcolU}
    \fME{0.5}{0.5}
    \fME{-0.5}{1.5}
     \node at (0.5,-.85) {\scalebox{.8}{$\phi_2$}};
    \node at (-0.5,-.85) {\scalebox{.8}{$\phi_1$}};
  \end{tikzpicture}=  \frac{1}{2}
  \begin{tikzpicture}[baseline={([yshift=-0.6ex]current bounding box.center)},scale=0.5]
    \prop{1}{1}{FcolU}
    \fME{1.5}{0.5}
    \fME{1.5}{1.5}
  \end{tikzpicture},
\end{equation}
{the fixed points of the transfer matrix $\mathbb T$ are of the form given in Fig.~\ref{fig:fixedpoints}a. To see this consider $\bra{L_{\mathrm{s}}}\mathbb T$, with $\bra{L_{\mathrm{s}}}$ given in Fig.~\ref{fig:fixedpoints}a. Starting from above we apply repeatedly the first of~\eqref{eq:LR} until we remove the leftmost column of gates. We then proceed with removing the second column up to the gate applied on the initial state. The latter can be removed using the first of~\eqref{eq:solvablestates} while the numerical factors combine to give $\bra{L_{\mathrm{s}}}\mathbb T= \bra{L_{\mathrm{s}}}$. Analogously, using the right relations of \eqref{eq:LR} and \eqref{eq:solvablestates} gives $\mathbb T\ket{R_{\mathrm{s}}}=\ket{R_{\mathrm{s}}}$. Here and in the following we add the subscript ``$\mathrm{s}$" to quantities computed for initial states fulfilling \eqref{eq:solvablestates}.}

Expressing a general qubit state as $\ket{\phi}= r e^{i \alpha}\ket{0}+\sqrt{1-r^2}e^{i \beta} \ket{1}$, a simple calculation reveals that all states $\ket{\phi_1}\otimes\ket{\phi_2}$ fulfilling \eqref{eq:solvablestates} can be parameterised as follows 
\be
\begin{aligned}
  0 &=
  r_1r_2\sqrt{1-r_1^2}\sqrt{1-r_2^2} \cos(\alpha_1\!-\!\beta_1)\cos(\alpha_2\!-\!\beta_2) ,
  \\
  0 &=
  \left (r^2_1-\frac{1}{2}\right)\left(r^2_2-\frac{1}{2}\right) ,
\end{aligned}
\ee
where the subscripts $1,2$ refer to parameters of $\ket{\phi_{1,2}}$. The first equation is fulfilled iff the first of~\eqref{eq:solvablestates} holds while the second iff the second of~\eqref{eq:solvablestates} holds. 

A remarkable property of the DFQE is that we do not need to fulfil both these conditions to have simple fixed points. If $\ket{\phi_{1,2}}$ are both classical configurations $\ket{0}$ or $\ket{1}$, only the first of \eqref{eq:solvablestates} is fulfilled, however the fixed points display the simple form of Fig.~\ref{fig:fixedpoints}b. Similarly, if both $\ket{\phi_{1,2}}$ are the flat superpositions $\ket{\phi_1}=\ket{\phi_2}=\ket{-}=(\ket{0}+\ket{1})/\sqrt{2}$, only the second of \eqref{eq:solvablestates} holds but the fixed points take the simple form in Fig.~\ref{fig:fixedpoints}c. This is because the local gate fulfils the following  relations     
\be
  \begin{tikzpicture}[baseline={([yshift=-0.6ex]current bounding box.center)},scale=0.5]
    \prop{0}{0}{FcolU}
    \prop{-1}{1}{FcolU}
   \MErd{-1.5}{0.5}
   \MErd{-0.5}{-0.5}
    \fME{-1.5}{1.5}
  \end{tikzpicture}=
  \begin{tikzpicture}[baseline={([yshift=-0.6ex]current bounding box.center)},scale=0.5]
    \prop{1}{0}{FcolU}
    \gridLine{0.5}{1.5}{1.5}{1.5}
    \fME{0.5}{0.5}
    \gridLine{.5}{1.5-0.25}{.5}{1.5+0.25}
    \MErd{0.5}{-.5}
  \end{tikzpicture},
  \qquad 
    \begin{tikzpicture}[baseline={([yshift=-0.6ex]current bounding box.center)},scale=0.5]
    \prop{0}{0}{FcolU}
    \prop{1}{1}{FcolU}
    \node at (1.75,.25) {\scalebox{0.85}{${\underline s_2}$}};
    \node at (.75,.25-1) {\scalebox{0.85}{${\underline s_1}$}};
    \fME{1.5}{1.5}
  \end{tikzpicture}=
  \begin{tikzpicture}[baseline={([yshift=-0.6ex]current bounding box.center)},scale=0.5]
    \prop{1}{0}{FcolU}
    \gridLine{0.5}{1.5}{1.5}{1.5}
    \fME{1.5}{0.5}
    \node at (2.5,1.5) {\scalebox{0.85}{$\,\,\underline s_1\!+\! \underline s_2$}};
    \node at (1.75,.25-1) {\scalebox{0.85}{${\underline s_1}$}};
  \end{tikzpicture},
  \label{eq:additionalconditions}
\ee
where we have introduced the diagrams and notation
\be
\!\!\!\!\begin{tikzpicture}[baseline={([yshift=-0.6ex]current bounding box.center)},scale=0.5]
    \gridLine{0}{0}{0}{.5}
    \MEh{0}{0}
  \end{tikzpicture}=\ket{-}\otimes_r \ket{-}, \,
  \begin{tikzpicture}[baseline={([yshift=-0.6ex]current bounding box.center)},scale=0.5]
   \gridLine{0}{-0.25}{0}{.25}
     \node at (0,-.5) {\scalebox{0.85}{${\underline s}$}};
  \end{tikzpicture}=\ket{s\, {\rm mod}\, 2}\otimes_r \ket{s\, {\rm mod}\, 2}, \, s\in \mathbb N_{0}\, ,
  \label{eq:sbar}
 \ee  
{ and} $\otimes_r$ indicates that the tensor product is between the states of the same site in the forward and backward branches, see Fig.~\ref{fig:circuit}a. { The simplification mechanism is very similar to the one discussed after Eq.~\eqref{eq:solvablestates} and we refer the reader to the Supplemental Material \cite{Note10}. In the following we add the subscripts ``$\mathrm{cl}$" and ``${\mathrm{F}}$" to quantities computed respectively for initial states that are classical configurations and flat superpositions of them \footnote{
Even though the space transfer matrix depends on the initial state, in order to simplify the notation,  we will use the same symbol $\mathbb T$ to denote the space transfer matrix of the three classes of states}.
}  


\begin{figure*}[t]
\includegraphics[width=1.0\textwidth]{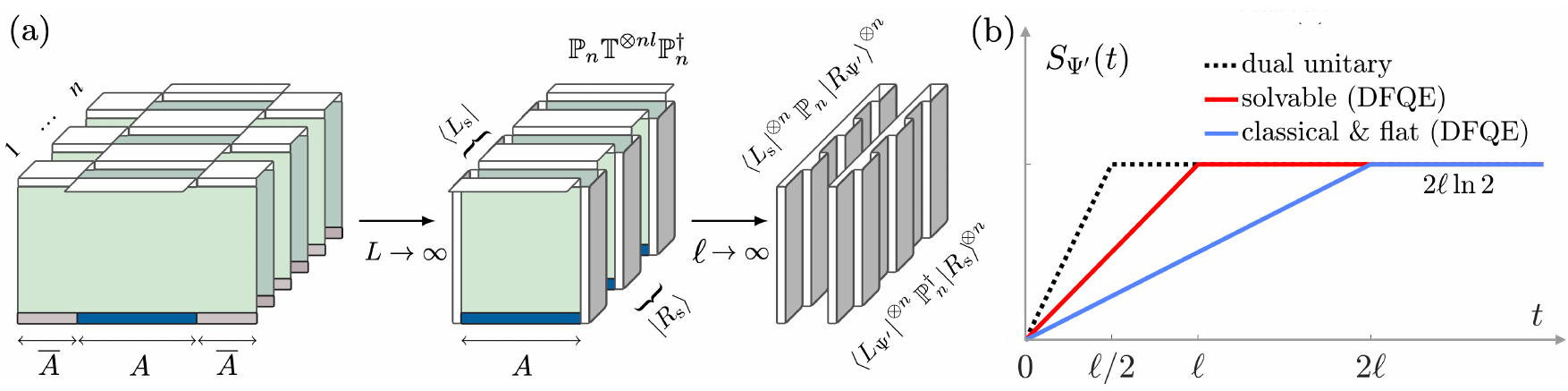}
  \caption{ 
    {R\'enyi entropies and entanglement dynamics.}
    (a) Diagrammatic representation of $\tr[\rho_{A}^n]$ for $n=3$. (b) Comparison of entanglement growth between DFQE model (solid lines) and dual-unitary circuits (dashed line). For the DFQE model, the system $A$ of length $\ell \ll L$ is initially prepared in a {solvable} state $|\Psi^\prime \rangle$ fulfilling \eqref{eq:solvablestates} and \eqref{eq:solvableSubsetMT} (red line),
    or one of  \eqref{eq:solvablestates} and one of \eqref{eq:additionalconditions} (blue line). For $t> 2\ell$ all curves reach the infinite-temperature value $2\ell \ln 2$.
    }
  \label{fig:Fig3}
\end{figure*}

\noindent
{\textit{Subsystem Dynamics.---}}  
The exact expressions for the fixed points in Fig.~\ref{fig:fixedpoints} represent our first main result. To illustrate their power, we use them to determine the relaxation time of a block of $2\ell$ qubits in a region denoted by $A$. In particular, we focus on the quench from the state $\ket{\Psi_0}=(\ket{\phi_1}\!\otimes\!\ket{\phi_2})^{\otimes L - \ell}\!\otimes\!\ket{\Psi'}$, where $\ket{\phi_1}\!\otimes\!\ket{\phi_2}$ are solvable states \eqref{eq:solvablestates} and $\ket{\Psi'}$ is arbitrary~\footnote{Since we took periodic boundary conditions, the actual position of the block is irrelevant.}. Using the exact expressions in Fig.~\ref{fig:fixedpoints}a we find that the reduced density matrix at time $t$ can be simplified as follows
\begin{equation}
    \label{eq:densitymatrix}
  \rho_{A,\Psi^{\prime}}(t) = \frac{1}{2^{2t}}
  \begin{tikzpicture}[baseline={([yshift=-0.6ex]current bounding box.center)},scale=0.5]
    \foreach \y in {0,2,...,6}{
      \foreach\x in {0,2,4,6}{\prop{\x}{\y+1}{FcolU}}
      \foreach\x in {0,2,4}{\prop{1+\x}{\y}{FcolU}}
      \fME{-0.5}{\y+1.5}
      \fME{-0.5}{\y+0.5}
      \fME{6.5}{\y+1.5}
      \fME{6.5}{\y+0.5}}
      \draw[rounded corners=1, fill = colLines]  (0,-0.5) rectangle (6,-.75);
      \node at (7.25,-0.75) {\scalebox{.8}{$\ket{\Psi^{\prime}}\!\otimes\!\ket{\Psi^{\prime}}$}};
    \end{tikzpicture}.
\end{equation}
Note how Eq.~\eqref{eq:densitymatrix} does not contain explicitly the environment ($\bar A$, the complement of $A$){: its effect is encoded in the boundaries of the time-evolution operator of $A$. In other words, the superoperator formed by two subsequent horizontal layers of the tensor network in Eq.~\eqref{eq:densitymatrix},
\be
  \mathcal{C}_x=\frac{1}{4}
  \begin{tikzpicture}[baseline={([yshift=-0.6ex]current bounding box.center)},scale=0.5]
    \foreach\x in {0,2,4,6}{\prop{\x}{1}{FcolU}}
    \foreach\x in {0,2,4}{\prop{1+\x}{0}{FcolU}}
    \fME{-0.5}{1.5}
    \fME{-0.5}{0.5}
    \fME{6.5}{0.5}
    \fME{6.5}{1.5}
    \draw[|<->|] (0,-1) -- (6,-1) node[midway,below] {\scalebox{0.7}{$2x$}};
  \end{tikzpicture}.
\ee
retains information on $\bar A$ only through its depolarising boundaries.} In general, tracing out part of a unitarily evolving system gives rise to non-Markovian dissipative evolution on the subsystem \cite{gardiner2004quantum}. In constrast, for the case of the DFQE the evolution of the subsystem is Markovian 
and the superoperator $\mathcal{C}_x$ is a time-local quantum map. Equation~\eqref{eq:densitymatrix} represents a drastic simplification: the dynamics of a block of $2\ell$ spins can be fully determined by diagonalising a $4^{2\ell}\times 4^{2\ell}$ matrix, which can be done analytically for small $\ell$ and numerically larger $\ell$. Moreover, one can use Eq.~\eqref{eq:densitymatrix} to show \cite{Note10} that $\rho_{A}(t)=\Id/2^{2\ell}$ for any $t \geq 2 \ell$, so that the subsystem reaches the maximal entropy state in a \emph{finite} number of steps. {This result contrasts with what happens in generic systems, where the presence of exponential corrections means that the stationary state is only reached exactly at infinite time. An analogous situation to the one here is found in dual-unitary circuits~\cite{kos2021thermalization}, with an important difference: in the DFQE the number of steps to approach stationarity is twice larger than for dual-unitaries.}

\noindent
{\textit{Entanglement.---}}
Using the above properties we can compute the growth of entanglement from various homogeneous and inhomogeneous initial states. For concreteness we consider a system that is prepared in a solvable state everywhere, except for a finite subsystem $A$ of length $\ell=|A|$. At some later time $t$ the R\'enyi entanglement entropy between $A$ and the rest is defined as 
\begin{equation}
  S_{\Psi^{\prime}}^{(n)}(t)=\frac{1}{1-n}\tr[\rho_{A,\Psi^{\prime}}^n(t)],
  \label{eq:renyi}
\end{equation}
where $n$ is the R\'enyi index, and $\rho_{A,\Psi^{\prime}}(t)$ is given in Eq.~\eqref{eq:densitymatrix}. Within $A$ the system is prepared in one of the classes of solvable states: those fulfilling both  Eqs.~\eqref{eq:solvablestates}, homogeneous \emph{flat} states, or classical configurations.

The entanglement entropies show different scalings depending on the ratio between the subsystem size $\ell$, and time $t$. In particular, we expect a simple result in the limit $t\to\infty$, where the entropies saturate at a value extensive in $\ell$ (see Fig.~\ref{fig:Fig3}b). In fact, the finite-time relaxation discussed above implies that for {\em any initial state} in $A$ and $t\ge 2\ell$ all R\'enyi entropies are the same
\begin{equation}
  \left.S_{\Psi^{\prime}}^{(n)}(t)\right|_{t>2\ell}= 2\ell \ln 2 .
\end{equation}
{For times that are shorter than $2\ell$, there are no immediate simplifications at the level of the single reduced density matrix, but we need to consider the full trace in Eq.~\eqref{eq:renyi}. 
Assuming for definiteness that inside of the subsystem $A$ the initial state $\Psi^{\prime}$ is a product state and is invariant under lattice shifts by an even number of sites, the generalised purity can be compactly expressed in terms of the corresponding space transfer-matrix $\mathbb{T}
$ as}
\begin{equation}
  \tr[\rho_{A,\Psi^{\prime}}^n(t)]=
  \bra{L_\mathrm{s}}^{\otimes n}
  \mathbb{P}_n
  \mathbb{T}^{\otimes n\,|A|}
  \mathbb{P}_n^{\dagger}
  \ket{R_\mathrm{s}}^{\otimes n},
  \label{eq:tr}
\end{equation}
where $\mathbb{P}_n$ is an operator that implements the permutation of $n$
copies, see Fig.~\ref{fig:Fig3}a. This expression suggests another conceptually simple regime: the ``early time" regime where $t$ is fixed and $|A|$ is large. In this regime Eq.~\eqref{eq:tr} is written in terms of a large power of a finite matrix 
{expressible in terms of a fixed point: whenever $t<\ell/2$ the powers of the transfer matrix factorize}
\begin{equation}
  \left.\mathbb{T}^{x}\right|_{x>2t}
  =\ketbra{R_{\Psi^{\prime}}}{L_{\Psi^{\prime}}},
\end{equation}
and \eqref{eq:tr} reduces to a product of two matrix elements,
\begin{equation}
 \label{eq:reducedEarlyTimeAll}
  \tr[\rho_{A,\Psi^{\prime}}^n(t)]\mkern-4mu=\mkern-4mu
  \bra{L_{\mathrm{s}}}^{\otimes n} \mathbb{P}_n
  \ket{R_{\Psi^{\prime}}}^{\mkern-2mu\otimes n}
  \mkern-2mu
  \bra{L_{\Psi^{\prime}}}^{\otimes n} \mathbb{P}_n^{\dagger}
  \ket{R_{\mathrm{s}}}^{\mkern-2mu\otimes n}\mkern-6mu.
\end{equation}
With the specific form of fixed points, Fig.~\ref{fig:fixedpoints}, we can evaluate these overlaps and obtain for the three classes~\cite{Note10}
\begin{equation}\label{eq:entropyEarlyTime}
  \begin{gathered}
  S_{\mathrm{s}}^{(n)}(t)|_{2t<\ell}=2t\ln 2,\\
  S_{\mathrm{cl}}^{(n)}(t)|_{2t<\ell}=
  S_{\mathrm{F}}^{(n)}(t)|_{t<\ell}=t\ln 2.
  \end{gathered}
\end{equation}
{Interestingly, we see that the entanglement entropies all grow with the same slope. Moreover, for classical configurations and their flat superposition this slope is reduced by 2: this is an explicit example of the initial-state dependence of the entanglement velocity. Note that for the flat superposition the range of validity of the early time expression is larger ($t<\ell$ rather than $2t<\ell$) due to the flat state being locally invariant under the dynamics.}

The hardest regime to access is the intermediate-time regime $\ell/2<t<2\ell$. In this case Eq.~\eqref{eq:tr} does not directly factorise, and cannot be determined only knowing the fixed points. Remarkably, and in contrast to other known solvable examples~\cite{klobas2021exact,klobas2021entanglement}, Eq.~\eqref{eq:tr} can be evaluated also in this regime for the three cases considered here. This leads to a complete description of the entanglement dynamics so far only attained for dual-unitary circuits~\cite{bertini2019exact, piroli2020exact}. In particular, when the subsystem is prepared in a classical configuration, or the flat state, the partition sum~\eqref{eq:tr} evaluates to $2^{(1-n)t}$ for all $t$ in the intermediate regime, i.e., $\ell/2<t<2\ell$. This gives~\cite{Note10}
    \begin{equation}
    \label{eq:fullentropyclassicalflat}
      S_{\mathrm{cl}}^{(n)}(t)=S_{\mathrm{F}}^{(n)}(t)=
      \min(t,2\ell) \ln 2.
    \end{equation}
    Instead, if the subsystem is prepared in a \emph{solvable} state that additionally satisfies
    \begin{equation}\label{eq:solvableSubsetMT}
      \begin{tikzpicture}[baseline={([yshift=-0.6ex]current bounding box.center)},scale=0.5]
        \gridLine{0}{0}{-0.75}{0.75}
        \prop{0}{0}{FcolU}
        \Pd{-0.5}{0.5}
        \node at (-0.75,-0.75) {\scalebox{0.85}{$\phi_1$}};
        \node at (0.75,-0.75) {\scalebox{0.85}{$\phi_2$}};
      \end{tikzpicture}
      =
      \begin{tikzpicture}[baseline={([yshift=-0.6ex]current bounding box.center)},scale=0.5]
        \gridLine{0}{0}{-0.75}{0.75}
        \gridLine{0}{0}{0.75}{0.75}
        \prop{0}{0}{FcolU}
        \Pd{0.5}{0.5}
        \Pd{-0.5}{0.5}
        \node at (-0.75,-0.75) {\scalebox{0.85}{$\phi_1$}};
        \node at (0.75,-0.75) {\scalebox{0.85}{$\phi_2$}};
      \end{tikzpicture},\qquad
      \begin{tikzpicture}[baseline={([yshift=-0.6ex]current bounding box.center)},scale=0.5]
        \gridLine{0}{-0.5}{0}{0.5}
        \Pd{0}{0}
      \end{tikzpicture}=
      \begin{bmatrix}
        1 & 0 & 0 & 0 \\
        0 & 0 & 0 & 0 \\
        0 & 0 & 0 & 0 \\
        0 & 0 & 0 & 1
      \end{bmatrix},
  \end{equation}
    the stationary state is reached at $t=\ell$ (rather than $2\ell$), and for ${\ell/2<t<\ell}$
    Eq.~\eqref{eq:tr} gives $2^{2t(1-n)}$. This implies
    \begin{equation}
      S_{\mathrm{s}}(t)=\min(2t,2\ell) \ln 2.
      \label{eq:fullentropysolvable}
    \end{equation}
{An interesting question is whether this surprising result can be ascribed to a general property of the initial states. While the DFQE is a Clifford circuit 
the above is not a consequence of initial states being stabilizer states: even though classical configurations and the flat state are stabilizers, the solvable states fulfilling Eqs.~\eqref{eq:solvablestates} and \eqref{eq:solvableSubsetMT} are generically not. As shown in \cite{Note10}, the exact results in Eqs.~\eqref{eq:fullentropyclassicalflat} and \eqref{eq:fullentropysolvable} rely on three different ``microscopic mechanisms'' of simplifications that combine properties of the initial states and the time-evolution. The overall effect, however, is similar in the three cases: the microscopic simplifications decouple the entanglement production at the two boundaries between $A$ and $\bar A$. This allows to treat the problem as if it were always in the early time regime. Importantly, Eq.~\eqref{eq:solvableSubsetMT} is a necessary requirement for this to happen: finite-time numerics show that for solvable states that do not satisfy that condition the entanglement entropy deviates from \eqref{eq:fullentropysolvable} at intermediate times.}

\noindent
{\textit{Conclusions.---}}
We have solved exactly the entanglement dynamics of the deterministic Floquet quantum East model, a quantum circuit defined in terms of local CNOT gates that implement the same kinetic constraint as the East model \cite{jackle1991a-hierarchically,horssen2015dynamics}. To our knowledge, these are the first exact results for entangling dynamics in an interacting non-integrable circuit beyond those in the dual-unitary class. The simplicity of the DFQE model allows its dynamics to be solved in the large size limit for a broad class of initial product states that extends beyond Clifford stabilizers, exploiting the techniques of propagation-in-space. One can think of many avenues for future research. An immediate one is to characterise exactly operator spreading by extending the results of Ref.~\cite{foligno2023quantum} on the butterfly velocity of DU2 circuits to determine the full profile of out-of-time-ordered correlators. Other directions include studying the effect of local measurements on entanglement at the level of quantum trajectories~\cite{li2018quantum,li2019measurement,skinner2019measurement,chan2019unitary} and the quantification of dynamical fluctuations as is done in the classical Floquet East~\cite{klobas2023exact}. \\

\begin{acknowledgments}
We acknowledge financial support from the Royal Society through the University Research Fellowship No.\ 201101 (B.\ B.), from EPSRC Grant No.\ EP/V031201/1 (C.\ D.\ F.\ and J.\ P.\ G.), and from The Leverhulme Trust through the Early Career Fellowship No.\ ECF-2022-324 (K.\ K.). B.\ B.\ and  K.\ K.\ warmly acknowledge the hospitality of the Simons Center for Geometry and Physics during the program ``Fluctuations, Entanglements, and Chaos: Exact Results'' where this work was completed.
\end{acknowledgments}

\footnotetext[10]{See the Supplemental Material.}

\bibliographystyle{apsrev4-2}
\bibliography{bibliography}

\onecolumngrid
\begin{center}
  {\Large \bf Supplemental Material}
\end{center}
\onecolumngrid
\newcounter{equationSM}
\newcounter{figureSM}
\newcounter{tableSM}
\stepcounter{equationSM}
\setcounter{equation}{0}
\setcounter{figure}{0}
\setcounter{table}{0}
\setcounter{section}{0}
\makeatletter
\renewcommand{\theequation}{\textsc{sm}-\arabic{equation}}
\renewcommand{\thefigure}{\textsc{sm}-\arabic{figure}}
\renewcommand{\thetable}{\textsc{sm}-\arabic{table}}

Here we report some useful information complementing the main text. In particular
\begin{itemize}
  \item[-] In Sec.~\ref{sec1} we introduce the space transfer matrix and compute the corresponding fixed points for three choices of initial states.
  \item[-] In Secs.~\ref{sec2} we provide additional details on the computation of the entanglement entropy for three distinct time regimes.
\end{itemize}
\section{Exact fixed points} \label{sec1}

We define the transfer matrix in space  $\mathbb T$ at time $t$ for a given two-site state $|\phi_1 \rangle \otimes | \phi_2 \rangle$ as:
\begin{equation}
\mathbb T= 
 \begin{tikzpicture}[baseline={([yshift=-0.6ex]current bounding box.center)},scale=0.5]
    \foreach \y in {0,2,...,8}{
      \foreach\x in {0}{\prop{\x}{\y+1}{FcolU}}
      \foreach\x in {0}{\prop{1+\x}{\y}{FcolU}}
      }
      \fME{1-0.5}{8+1.5}
      \fME{1.60-0.5}{8+1.5}
       \node at (1-0.75,-0.9) {\scalebox{0.85}{$\phi_1$}};
       \node at (1+0.75,-0.9) {\scalebox{0.85}{$\phi_2$}};
       \gridLine{1.5+0}{9+0}{1.5-0.40}{9+0.40}
     \draw[|<->|] (2,0.5) -- (2,9) node[midway,right] {\scalebox{0.7}{$2t$}};
    \end{tikzpicture}\,,
\end{equation}
the corresponding fixed points, i.e., left- and right-eigenvectors with largest eigenvalue $\lambda=1$, can be computed for the three classes of states presented in the letter (results are shown in Fig.\ref{fig:fixedpoints}). In this section we want to prove they satisfy the fixed point relations for each class of states. \\

We first consider the states  $|\phi_1 \rangle $ and $|\phi_2 \rangle$ fulfilling both relations of Eq.~\eqref{eq:solvablestates}. The corresponding (normalised) fixed points are 
\begin{equation} \label{eq:fixedpts1}
    \langle L_{\mathrm{s}} | = 2^{-t+1}\, \begin{tikzpicture}[baseline={([yshift=-0.6ex]current bounding box.center)},scale=0.5]
    \foreach \y in {0,2,...,8}{
      \foreach\x in {0}{\prop{1+\x}{\y}{FcolU}}
      }
       \foreach \y in {0,2,...,6}{
      \fME{0.50}{\y+1.5}
      \fME{0.50}{\y+0.5}
      }
      
      \fME{0.5}{8+0.5}
       \node at (1-0.75,-0.9) {\scalebox{0.85}{$\phi_1$}};
       \node at (1+0.75,-0.9) {\scalebox{0.85}{$\phi_2$}};
        \fME{1.5-0.5}{8+1.5}
        \gridLine{1.5+0}{9+0}{1.5-0.40}{9+0.40}
    \end{tikzpicture}\,, \quad
    | R_{\mathrm{s}} \rangle =  2^{-t}\, \begin{tikzpicture}[baseline={([yshift=-0.6ex]current bounding box.center)},scale=0.5]
   \foreach \y in {0,2,...,8}{
      \foreach\x in {0}{\prop{\x}{\y+1}{FcolU}}
      }
        \foreach \y in {0,2,...,8}{
      \fME{0.5}{\y+1.5}
      \fME{0.5}{\y+0.5}
      }
      \end{tikzpicture}\,,
\end{equation}
where the normalisation is chosen such that $\langle L_{\mathrm{s}} | R_{\mathrm{s}} \rangle =1$. The left vector indeed satisfies
\begin{equation}
\label{eq:Lvec}
    \bra{L_{\mathrm{s}}} \mathbb T=  2^{-t+1}\!
 \begin{tikzpicture}[baseline={([yshift=-0.6ex]current bounding box.center)},scale=0.5]
    \foreach \y in {0,2,...,8}{
      \foreach\x in {0}{\prop{\x}{\y+1}{FcolU}}
      \foreach\x in {-2,0}{\prop{1+\x}{\y}{FcolU}}
      }
      \fME{-0.5}{8+1.5}
      \fME{1-0.5}{8+1.5}
      \fME{1.5-0.5}{8+1.5}
       \node at (-1-0.55,-0.9) {\scalebox{0.85}{$\phi_1$}};
       \node at (-1+0.55,-0.9) {\scalebox{0.85}{$\phi_2$}};
       \node at (1-0.55,-0.9) {\scalebox{0.85}{$\phi_1$}};
       \node at (1+0.55,-0.9) {\scalebox{0.85}{$\phi_2$}};
       \foreach \y in {0,2,...,6}{
      \fME{-1-0.5}{\y+1.5}
      \fME{-1-0.5}{\y+0.5}
      }
       \fME{-1-0.5}{8.5}
            \gridLine{1.5+0}{9+0}{1.5-0.40}{9+0.40}
    \end{tikzpicture}
    \mkern-14mu=  2^{-t+2}\!
    \begin{tikzpicture}[baseline={([yshift=-0.6ex]current bounding box.center)},scale=0.5]
    \foreach \y in {0,2,...,6}{
      \foreach\x in {0}{\prop{\x}{\y+1}{FcolU}}
      \foreach\x in {-2,0}{\prop{1+\x}{\y}{FcolU}}
      }
      \foreach\x in {0}{\prop{1+\x}{8}{FcolU}}

      \fME{-0.5}{6+1.5}
      \fME{1-0.5}{7+1.5}
      \fME{1.5-0.5}{8+1.5}
       \node at (-1-0.55,-0.9) {\scalebox{0.85}{$\phi_1$}};
       \node at (-1+0.55,-0.9) {\scalebox{0.85}{$\phi_2$}};
       \node at (1-0.55,-0.9) {\scalebox{0.85}{$\phi_1$}};
       \node at (1+0.55,-0.9) {\scalebox{0.85}{$\phi_2$}};
       \foreach \y in {0,2,...,4}{
      \fME{-1-0.5}{\y+1.5}
      \fME{-1-0.5}{\y+0.5}
      }
       \fME{-1-0.5}{6.5}
            \gridLine{1.5+0}{9+0}{1.5-0.40}{9+0.40}
    \end{tikzpicture}
    \mkern-14mu=  2^{-t+2}\mkern-24mu
    \begin{tikzpicture}[baseline={([yshift=-0.6ex]current bounding box.center)},scale=0.5]
    \foreach \y in {0,2,...,6}{
      \foreach\x in {0}{\prop{\x}{\y+1}{FcolU}}
      \foreach\x in {0}{\prop{1+\x}{\y}{FcolU}}
      }
      \foreach\x in {-2}{\prop{\x+1}{0}{FcolU}}
       \foreach\x in {0}{\prop{1+\x}{8}{FcolU}}

      \fME{-0.5}{6+1.5}
      \fME{1-0.5}{7+1.5}
      \fME{1.5-0.5}{8+1.5}
       \node at (-1-0.55,-0.9) {\scalebox{0.85}{$\phi_1$}};
       \node at (-1+0.55,-0.9) {\scalebox{0.85}{$\phi_2$}};
       \node at (1-0.55,-0.9) {\scalebox{0.85}{$\phi_1$}};
       \node at (1+0.55,-0.9) {\scalebox{0.85}{$\phi_2$}};
       \foreach \y in {2}{
      \fME{-0.5}{\y+1.5}
      \fME{-0.5}{\y+0.5}
      }
      \fME{-1-0.5}{0.5}
       \fME{-0.5}{1.5}
       \fME{-0.5}{4.5}
       \fME{-0.5}{5.5}
       \fME{-0.5}{6.5}
            \gridLine{1.5+0}{9+0}{1.5-0.40}{9+0.40}
    \end{tikzpicture}
    \mkern-14mu=  2^{-t+1}
    \begin{tikzpicture}[baseline={([yshift=-0.6ex]current bounding box.center)},scale=0.5]
    \foreach \y in {0,2,...,6}{
      \foreach\x in {0}{\prop{\x}{\y+1}{FcolU}}
      \foreach\x in {0}{\prop{1+\x}{\y}{FcolU}}
      }
       \foreach\x in {0}{\prop{1+\x}{8}{FcolU}}

      \fME{-0.5}{6+1.5}
      \fME{1-0.5}{7+1.5}
      \fME{1.5-0.5}{8+1.5}
       \node at (1-0.55,-0.9) {\scalebox{0.85}{$\phi_1$}};
       \node at (1+0.55,-0.9) {\scalebox{0.85}{$\phi_2$}};
       \foreach \y in {0,2}{
      \fME{-0.5}{\y+1.5}
      \fME{-0.5}{\y+0.5}
      }
       \fME{-0.5}{4.5}
       \fME{-0.5}{5.5}
       \fME{-0.5}{6.5}
            \gridLine{1.5+0}{9+0}{1.5-0.40}{9+0.40}
    \end{tikzpicture}
    \mkern-14mu=  2^{-t+1} \!
    \begin{tikzpicture}[baseline={([yshift=-0.6ex]current bounding box.center)},scale=0.5]
    \foreach \y in {0,2,...,8}{
      \foreach\x in {0}{\prop{1+\x}{\y}{FcolU}}
      }
       \foreach \y in {0,2,...,6}{
      \fME{0.5}{\y+1.5}
      \fME{0.5}{\y+0.5}
      }
      
      \fME{0.5}{8+0.5}
       \node at (1-0.75,-0.9) {\scalebox{0.85}{$\phi_1$}};
       \node at (1+0.75,-0.9) {\scalebox{0.85}{$\phi_2$}};
        \fME{1.5-0.5}{8+1.5}
        \gridLine{1.5+0}{9+0}{1.5-0.40}{9+0.40}
    \end{tikzpicture} \mkern-14mu= \langle L_{\mathrm{s}} | \,,
\end{equation}
where we start contracting the above tensor network from the top by using 
\begin{equation}
      \begin{tikzpicture}[baseline={([yshift=-0.6ex]current bounding box.center)},scale=0.5]
    \prop{0}{0}{FcolU}
    \prop{1}{1}{FcolU}
    \fME{-0.5}{0.5}
    \fME{0.5}{1.5}
    \fME{1.5}{1.5}
    \fME{-0.5}{-0.5}
  \end{tikzpicture}= 2 \; \begin{tikzpicture}[baseline={([yshift=-0.6ex]current bounding box.center)},scale=0.5]
    \gridLine{0.5}{-0.25}{1.5}{-0.25}
     \gridLine{0.5}{0.50}{1.5}{0.50}
    \fME{0.5}{0.5}
    \fME{0.5}{-0.25}
  \end{tikzpicture}\,.
\end{equation}

The leftmost column is then removed by repeatedly applying the left of \eqref{eq:LR}, and the left of \eqref{eq:solvablestates} in the end. In the second last equality we further apply the left of \eqref{eq:LR}, to contract the second column. \\

Similarly, the right vector $\ket{R_{\mathrm{s}}}$ satisfies:
\begin{equation} \label{eq:Rvec}
 \mathbb T | R_{\mathrm{s}} \rangle =  2^{-t}\,
    \begin{tikzpicture}[baseline={([yshift=-0.6ex]current bounding box.center)},scale=0.5]
    \foreach \y in {0,2,...,8}{
      \foreach\x in {0,2}{\prop{\x}{\y+1}{FcolU}}
      \foreach\x in {0}{\prop{1+\x}{\y}{FcolU}}
      }
      \fME{1.5}{8+1.5}
      \fME{1-0.5}{8+1.5}
       \node at (1-0.55,-0.9) {\scalebox{0.85}{$\phi_1$}};
       \node at (1+0.55,-0.9) {\scalebox{0.85}{$\phi_2$}};
       \foreach \y in {0,2,...,8}{
      \fME{2.5}{\y+1.5}
      \fME{2.5}{\y+0.5}
      }
    \end{tikzpicture}
    =  2^{-t} \,
     \begin{tikzpicture}[baseline={([yshift=-0.6ex]current bounding box.center)},scale=0.5]
    \foreach \y in {0,2,...,8}{
      \foreach\x in {0}{\prop{\x}{\y+1}{FcolU}}
      \foreach\x in {0}{\prop{1+\x}{\y}{FcolU}}
      }
        \foreach \y in {0,2,...,6}{
      \foreach\x in {2}{\prop{\x}{\y+1}{FcolU}}
      }
      \fME{1.5}{7+1.5}
      \fME{1-0.5}{8+1.5}
      \bendLud{2.5}{8.5}{9.5}
       \node at (1-0.55,-0.9) {\scalebox{0.85}{$\phi_1$}};
       \node at (1+0.55,-0.9) {\scalebox{0.85}{$\phi_2$}};
       \foreach \y in {0,2,...,6,8}{
      \fME{2.5}{\y+1.5}
      \fME{2.5}{\y+0.5}
      }
    \end{tikzpicture}
    = 2^{-t} \,2 \,
     \begin{tikzpicture}[baseline={([yshift=-0.6ex]current bounding box.center)},scale=0.5]
    \foreach \y in {0,2,...,8}{
      \foreach\x in {0}{\prop{\x}{\y+1}{FcolU}}
      \foreach\x in {0}{\prop{1+\x}{\y}{FcolU}}
      }
      \fME{1.5}{7+1.5}
      \fME{1-0.5}{8+1.5}
       \node at (1-0.55,-0.9) {\scalebox{0.85}{$\phi_1$}};
       \node at (1+0.55,-0.9) {\scalebox{0.85}{$\phi_2$}};
       \foreach \y in {0,2,...,6}{
      \fME{1.5}{\y+1.5}
      \fME{1.5}{\y+0.5}
      }
    \end{tikzpicture}
     =2^{-t} \, 2 \,
     \begin{tikzpicture}[baseline={([yshift=-0.6ex]current bounding box.center)},scale=0.5]
    \foreach \y in {0,2,...,8}{
      \foreach\x in {0}{\prop{\x}{\y+1}{FcolU}}
      }
      \foreach \y in {0}{
      \foreach\x in {0}{\prop{1+\x}{\y}{FcolU}}
      }
       \node at (1-0.55,-0.9) {\scalebox{0.85}{$\phi_1$}};
       \node at (1+0.55,-0.9) {\scalebox{0.85}{$\phi_2$}};
      \fME{0.5}{8+0.5}
      \fME{0.5}{8+1.5}
       \foreach \y in {2,4,...,6}{
      \fME{0.5}{\y+1.5}
      \fME{0.5}{\y+0.5}
      }
      \fME{0.5}{1.5}
      \fME{1.5}{0.5}
    \end{tikzpicture}
    =2^{-t} \,\begin{tikzpicture}[baseline={([yshift=-0.6ex]current bounding box.center)},scale=0.5]
   \foreach \y in {0,2,...,8}{
      \foreach\x in {0}{\prop{\x}{\y+1}{FcolU}}
      }
        \foreach \y in {0,2,...,8}{
      \fME{0.5}{\y+1.5}
      \fME{0.5}{\y+0.5}
      }
      \node at (0,-0.9) {\scalebox{0.85}{$\phantom{\phi_2}$}};
    \end{tikzpicture} = \ket{R_{\mathrm{s}}}\,.
\end{equation}

Above, the first contraction follows from unitarity and produces a factor $1/2$. We then contract the rightmost column iterating the right relation in \eqref{eq:solvablestates}. The same for the other column, where we also use the right of \eqref{eq:solvablestates} to contract the last bit, which cancels out the factor $1/2$. \\

We now consider the states $|\phi_1\rangle =|\phi_2\rangle = |-\rangle $. The corresponding (normalised) fixed points are:
\begin{equation} \label{eq:fixedpts2}
    \langle L_{\mathrm{F}} | =  \begin{tikzpicture}[baseline={([yshift=-0.6ex]current bounding box.center)},scale=0.5]
    \foreach \y in {0,2,...,8}{
      \foreach\x in {0}{\prop{1+\x}{\y}{FcolU}}
      }
       \foreach \y in {0,2,...,6}{
       \MErd{0.5}{\y+1.5}
      \fME{0.5}{\y+0.5}
      }
      \fME{0.5}{8+0.5}
      
      \MErd{0.5}{-0.50}
      \MEld{1+0.5}{-0.50}
        \fME{1.5-0.5}{8+1.5}
        \gridLine{1.5+0}{9+0}{1.5-0.40}{9+0.40}
    \end{tikzpicture}\,, \quad
    | R_{\mathrm{F}} \rangle = 2^{-t+1} \,\begin{tikzpicture}[baseline={([yshift=-0.6ex]current bounding box.center)},scale=0.5]
   \foreach \y in {0,2,...,8}{
      \foreach\x in {0}{\prop{\x}{\y+1}{FcolU}}
      }
        \foreach \y in {0,2,...,8}{
      \fME{0.5}{\y+1.5}
      \fME{0.5}{\y+0.5}
      }
      \end{tikzpicture}\,.
\end{equation}
Indeed the left vector satisfies
\begin{equation}
     \langle L_{\mathrm{F}}| \, \mathbb T= 
    \begin{tikzpicture}[baseline={([yshift=-0.6ex]current bounding box.center)},scale=0.5]
    \foreach \y in {0,2,...,8}{
      \foreach\x in {0}{\prop{\x}{\y+1}{FcolU}}
      \foreach\x in {-2,0}{\prop{1+\x}{\y}{FcolU}}
      }
      \fME{-0.5}{8+1.5}
      \fME{1-0.5}{8+1.5}
      \fME{1.5-0.5}{8+1.5}
       \MErd{-1-0.5}{-0.50}
      \MEld{-1+0.5}{-0.50}
       \MErd{1-0.5}{-0.50}
      \MEld{1+0.5}{-0.50}
       \foreach \y in {0,2,...,6}{
       \MErd{-1.5}{\y+1.5}
      \fME{-1.5}{\y+0.5}
      }
       \fME{-1-0.5}{8.5}
            \gridLine{1.5+0}{9+0}{1.5-0.40}{9+0.40}
    \end{tikzpicture}= 
    \begin{tikzpicture}[baseline={([yshift=-0.6ex]current bounding box.center)},scale=0.5]
 \foreach \y in {0}{
      \foreach\x in {0}{\prop{\x}{\y+1}{FcolU}}
      \foreach\x in {0}{\prop{1+\x}{\y}{FcolU}}
      }
       \foreach \y in {2,4,...,8}{
      \foreach\x in {0}{\prop{\x}{\y+1}{FcolU}}
      \foreach\x in {-2,0}{\prop{1+\x}{\y}{FcolU}}
      }
      \fME{-0.5}{8+1.5}
      \fME{1-0.5}{8+1.5}
      \fME{1.5-0.5}{8+1.5}
       \MErd{-1-0.5}{1.50}
      \MErd{-1+0.5}{+0.50}
       \MErd{1-0.5}{-0.50}
      \MEld{1+0.5}{-0.50}
       \foreach \y in {2,4,...,6}{
       \MErd{-1.5}{\y+1.5}
      \fME{-1.5}{\y+0.5}
      }
       \fME{-1-0.5}{8.5}
            \gridLine{1.5+0}{9+0}{1.5-0.40}{9+0.40}
    \end{tikzpicture}
    =  
    \begin{tikzpicture}[baseline={([yshift=-0.6ex]current bounding box.center)},scale=0.5]
    \foreach \y in {0,2,...,8}{
      \foreach\x in {0}{\prop{\x}{\y+1}{FcolU}}
      \foreach\x in {0}{\prop{1+\x}{\y}{FcolU}}
      }
      \fME{-0.5}{8+1.5}
      \fME{1-0.5}{8+1.5}
      \fME{1.5-0.5}{8+1.5}
      \MErd{1-0.5}{-0.50}
      \MEld{1+0.5}{-0.50}
       \foreach \y in {0,2,4,6}{
        \MErd{-0.5}{\y+0.5}
      \fME{-0.5}{\y+1.5}
      }
      \MErd{-0.5}{8.5}
     
            \gridLine{1.5+0}{9+0}{1.5-0.40}{9+0.40}
    \end{tikzpicture}
    =  
    \begin{tikzpicture}[baseline={([yshift=-0.6ex]current bounding box.center)},scale=0.5]
    \foreach \y in {0,2,...,8}{
      \foreach\x in {0}{\prop{1+\x}{\y}{FcolU}}
      }
       \foreach \y in {0,2,...,6}{
       \MErd{0.5}{\y+1.5}
      \fME{0.5}{\y+0.5}
      }
      \fME{0.5}{8+0.5}
      
      \MErd{0.5}{-0.50}
      \MEld{1+0.5}{-0.50}
        \fME{1.5-0.5}{8+1.5}
        \gridLine{1.5+0}{9+0}{1.5-0.40}{9+0.40}
    \end{tikzpicture} = \langle L_{\mathrm{F}} | \,.
\end{equation}
In this case, we use the left relation in~\eqref{eq:additionalconditions} to contract  both the left columns from the bottom to the top. The fact that $|R_{\mathrm{F}}\rangle$ satisfies $\mathbb{T} |R_{\mathrm{F}}\rangle =|R_{\mathrm{F}}\rangle$ follows immediately from \eqref{eq:Rvec} as the flat states fulfil the same relations used in the previous calculation. \\

We finally consider classical states i.e., $\ket{\phi_1} = \ket{s_1}$ and $\ket{\phi_2}=\ket{s_2}$, with $s_1,s_2\in\mathbb{Z}_2$. We have that the corresponding fixed points are:
\begin{equation} \label{eq:fixedpts3}
    \langle L_{\mathrm{cl}} | = 2^{-t+1} \,  \begin{tikzpicture}[baseline={([yshift=-0.6ex]current bounding box.center)},scale=0.5]
    \foreach \y in {0,2,...,8}{
      \foreach\x in {0}{\prop{1+\x}{\y}{FcolU}}
      }
       \foreach \y in {0,2,...,6}{
      \fME{0.5}{\y+1.5}
      \fME{0.5}{\y+0.5}
      }
      
      \fME{0.5}{8+0.5}
       \node at (1-0.75,-0.9) {\scalebox{0.85}{$\underline{s_1}$}};
       \node at (1+0.75,-0.9) {\scalebox{0.85}{$\underline{s_2}$}};
        \fME{1.5-0.5}{8+1.5}
        \gridLine{1.5+0}{9+0}{1.5-0.40}{9+0.40}
    \end{tikzpicture}\,, \quad    
    | R_{\mathrm{cl}} \rangle =\begin{tikzpicture}[baseline={([yshift=-0.6ex]current bounding box.center)},scale=0.5]
   \foreach \y in {0,2,...,8}{
      \foreach\x in {0}{\prop{\x}{\y+1}{FcolU}}
      }
        \foreach \y in {0,2,...,8}{
     \fME{0.5}{\y+1.5}
      }
       \node at (0.85,0.4) {\scalebox{0.85}{$\underline{s^\prime_1}$}};
       \node at (0.85,2.4) {\scalebox{0.85}{$\underline{s^\prime_2}$}};
       \node at (0.95,4.4) {\scalebox{0.85}{$\ldots$}};
       \node at (1.20,6.4) {\scalebox{0.85}{$\underline{s^\prime_{t-1}}$}};
       \node at (0.85,8.4) {\scalebox{0.85}{$\underline{s^\prime_{t}}$}};
      \end{tikzpicture}\,,
\end{equation}
where the classical configuration~$\lbrace s^\prime_j\rbrace$ is given as
\begin{equation}\label{eq:sprime}
  s^\prime_j \equiv s^\prime_{j-1}+s^\prime_{j-2}\pmod{2}\quad \text{for } \; 2< j\le t\,,
    \qquad 
    s^\prime_{2}={s_{2}}\,,\quad {s^\prime_{1}}\equiv {s_{1}+s_{2}}\pmod{2}\,.
\end{equation}
In particular $\lbrace s^\prime_j\rbrace$ is a periodic sequence with period $3$.
Since classical configurations fulfil the same relations as used in~\eqref{eq:Lvec}, the left vector is the same as in the solvable case, $\bra{L_{\mathrm{cl}}}=\bra{L_{\mathrm{s}}}$. To show that $\ket{R_{\mathrm{cl}}}$ as introduced above is the right fixed point, however, it requires some additional work. We start by applying the classical transfer matrix to $\ket{R_{\mathrm{cl}}}$, without specifying $s^{\prime}_j$ (this will be done at the end), and we contract the rightmost column by using the right of~\eqref{eq:additionalconditions},
\begin{equation}
 \mathbb{T}|  R_{\mathrm{cl}} \rangle =
     \begin{tikzpicture}[baseline={([yshift=-0.6ex]current bounding box.center)},scale=0.5]
    \foreach \y in {0,2,...,8}{
      \foreach\x in {0,2}{\prop{\x}{\y+1}{FcolU}}
      \foreach\x in {0}{\prop{1+\x}{\y}{FcolU}}
      }
      \fME{1.5}{8+1.5}
      \fME{1-0.5}{8+1.5}
       \node at (1-0.55,-0.9) {\scalebox{0.85}{$\underline{s_1}$}};
       \node at (1+0.55,-0.9) {\scalebox{0.85}{$\underline{s_2}$}};
        \foreach \y in {0,2,...,8}{
      \fME{2.5}{\y+1.5}
      }
     \node at (2+0.85,0.4) {\scalebox{0.85}{$\underline{s^\prime_1}$}};
       \node at (2+0.85,2.4) {\scalebox{0.85}{$\underline{s^\prime_2}$}};
       \node at (2+0.95,4.4) {\scalebox{0.85}{$\ldots$}};
       \node at (2+1.2,6.4) {\scalebox{0.85}{$\underline{s^\prime_{t-1}}$}};
       \node at (2+0.85,8.4) {\scalebox{0.85}{$\underline{s^\prime_{t}}$}};
    \end{tikzpicture}
    = 
     \begin{tikzpicture}[baseline={([yshift=-0.6ex]current bounding box.center)},scale=0.5]
    \foreach \y in {0,2,...,8}{
      \foreach\x in {0}{\prop{\x}{\y+1}{FcolU}}
      \foreach\x in {0}{\prop{1+\x}{\y}{FcolU}}
      }
        \foreach \y in {0,2,...,6}{
      \foreach\x in {2}{\prop{\x}{\y+1}{FcolU}}
      }
      \fME{1.5}{7+1.5}
      \fME{1-0.5}{8+1.5}
      \bendLud{2.5}{8.5}{9.5}
       \node at (1-0.55,-0.9) {\scalebox{0.85}{$\underline{s_1}$}};
       \node at (1+0.55,-0.9) {\scalebox{0.85}{$\underline{s_2}$}};
        \foreach \y in {0,2,...,8}{
      \fME{2.5}{\y+1.5}
      }
      \node at (2+0.85,0.4) {\scalebox{0.85}{$\underline{s^\prime_1}$}};
       \node at (2+0.85,2.4) {\scalebox{0.85}{$\underline{s^\prime_2}$}};
       \node at (2+0.95,4.4) {\scalebox{0.85}{$\ldots$}};
       \node at (2+1.2,6.4) {\scalebox{0.85}{$\underline{s^\prime_{t-1}}$}};
       \node at (2+0.75,8.4) {\scalebox{0.85}{$\underline{s^\prime_{t}}$}};
    \end{tikzpicture}=
    \begin{tikzpicture}[baseline={([yshift=-0.6ex]current bounding box.center)},scale=0.5]
    \foreach \y in {0,2,...,8}{
      \foreach\x in {0}{\prop{\x}{\y+1}{FcolU}}
      \foreach\x in {0}{\prop{1+\x}{\y}{FcolU}}
      }
      \fME{1-0.5}{8+1.5}
       \node at (1-0.55,-0.9) {\scalebox{0.85}{$\underline{s_1}$}};
       \node at (1+0.55,-0.9) {\scalebox{0.85}{$\underline{s_2}$}};
        \foreach \y in {0,2,...,8}{
      \fME{1.5}{\y+0.5}
      }
      \node at (1+0.85,1.4) {\scalebox{0.85}{$\underline{s^{\prime\prime}_2}$}};
       \node at (1+0.85,3.4) {\scalebox{0.85}{$\underline{s^{\prime\prime}_3}$}};
       \node at (1+0.95,5.4) {\scalebox{0.85}{$\ldots$}};
       \node at (1+0.85,7.4) {\scalebox{0.85}{$\underline{s^{\prime\prime}_{t}}$}};
    \end{tikzpicture}\,,
\end{equation}
where~\eqref{eq:additionalconditions} implies
\begin{equation}
  s^{\prime\prime}_{j} \equiv \sum_{k=1}^{j-1} s_k^{\prime} + s_2\pmod{2}.
\end{equation}
Repeating the same steps again we obtain
\begin{equation}
 \mathbb{T} |  R_{\mathrm{cl}} \rangle =
    \begin{tikzpicture}[baseline={([yshift=-0.6ex]current bounding box.center)},scale=0.5]
    \foreach \y in {0,2,...,8}{
      \foreach\x in {0}{\prop{\x}{\y+1}{FcolU}}
      }
    \foreach \y in {2,4,...,8}{
      \foreach\x in {0}{\prop{1+\x}{\y}{FcolU}}
      }
      \bendLud{1.5}{-0.5}{0.5}
      \fME{1-0.5}{8+1.5}
       \node at (1-0.25,1-0.8) {\scalebox{0.85}{$\underline{s^{\prime\prime\prime}_1}$}};
       \node at (1+0.55,-0.9) {\scalebox{0.85}{$\underline{s_2}$}};
        \foreach \y in {0,2,...,8}{
        \fME{1.5}{\y+0.5}
      }
      \node at (1+0.85,1.4) {\scalebox{0.85}{$\underline{s^{\prime\prime}_2}$}};
       \node at (1+0.85,3.4) {\scalebox{0.85}{$\underline{s^{\prime\prime}_3}$}};
       \node at (1+0.95,5.4) {\scalebox{0.85}{$\ldots$}};
       \node at (1+0.85,7.4) {\scalebox{0.85}{$\underline{s^{\prime\prime}_{t}}$}};
    \end{tikzpicture}
    =\begin{tikzpicture}[baseline={([yshift=-0.6ex]current bounding box.center)},scale=0.5]
   \foreach \y in {0,2,...,8}{
      \foreach\x in {0}{\prop{\x}{\y+1}{FcolU}}
      }
        \foreach \y in {0,2,...,8}{
      \fME{0.5}{\y+1.5}
      }
       \node at (0.9,0.4) {\scalebox{0.85}{$\underline{s^{\prime\prime\prime}_1}$}};
       \node at (0.9,2.4) {\scalebox{0.85}{$\underline{s^{\prime\prime\prime}_2}$}};
       \node at (0.95,4.4) {\scalebox{0.85}{$\ldots$}};
       \node at (1.1,6.4) {\scalebox{0.85}{$\underline{s^{\prime\prime\prime}_{t-1}}$}};
       \node at (0.9,8.4) {\scalebox{0.85}{$\underline{s^{\prime\prime\prime}_{t}}$}};
      \end{tikzpicture}\,,
      \qquad
      s_{1}^{\prime\prime\prime}\equiv s_1+s_2\pmod{2},\qquad
      s_{j}^{\prime\prime\prime}\equiv
      s_{1}^{\prime\prime\prime}+\sum_{k=2}^{j} s_{k}^{\prime\prime}\pmod{2}.
\end{equation}
Simplifying this we get
\begin{equation}
\label{eq:threeprimes}
    s^{\prime\prime\prime}_{2j} \equiv s_1+\sum_{k=1}^{j} s_{2k-1}^{\prime}\pmod{2},\qquad
    s^{\prime\prime\prime}_{2j-1} \equiv s_1+s_2+\sum_{k=1}^{j-1} s_{2k}^{\prime}\pmod{2}.
\end{equation}
Imposing now the fixed point condition (i.e., $s_j^{\prime\prime\prime}=s_j^{\prime}$) on~\eqref{eq:threeprimes}, we finally obtain the form in~\eqref{eq:sprime}.

\section{Entanglement Entropy}
\label{sec2}

\subsection{Early-time regime}  
Using the explicit form of fixed-points as given in Fig.~\ref{fig:fixedpoints}, we can 
express the factors appearing in the main-text Eq.~\eqref{eq:reducedEarlyTimeAll}
as a $2n\times 2t$ tensor network
\begin{equation}\label{eq:TNinitialregime}
  \bra{L_{\Psi_2}}^{\otimes n}\mathbb{P}_n\ket{R_{\mathrm{\Psi_1}}}^{\otimes n}=
  \bra{L_{\Psi_2}}^{\otimes n}\mathbb{P}_n^{\dagger}\ket{R_{\mathrm{\Psi_1}}}^{\otimes n}=
  \begin{tikzpicture}[baseline={([yshift=-0.6ex]current bounding box.center)},scale=0.6]
    \foreach \x in {0,2,4,6,8}{
      \draw[semithick,colLines,rounded corners=3.25]
      ({\x},7) -- ({\x+0.75},7.75) -- ({\x+1.25},7.75) -- ({\x+2},7);}
    \draw[ultra thick,white] (10,7) -- (9.5,7.5);

    \foreach \x in {0,2,4,6,8}{
      \propSt{\x+1}{0}{white}
      \props{\x}{1}{white}
      \node at ({\x+1},0) {\scalebox{0.7}{$C_{2}$}};
      \node at ({\x},1) {\scalebox{0.7}{$A_{1}$}};
      \foreach \y in {2,4,6}{
        \props{\x+1}{\y}{white}
        \props{\x}{\y+1}{white}
        \node at ({\x+1},{\y}) {\scalebox{0.7}{$B_2$}};
        \node at ({\x},{\y+1}) {\scalebox{0.7}{$A_{1}$}};
      }
    }
    \draw [very thick,gray,fill=gray, fill opacity=0.1,rounded corners=2] 
    (-0.75,6.5) rectangle (9.75,7.875);
    \draw [very thick,gray,fill=gray, fill opacity=0.1,rounded corners=2] 
    (-0.75,2.5) rectangle (9.75,4.5);
    \draw [very thick,gray,fill=gray, fill opacity=0.1,rounded corners=2] 
    (-0.75,-0.5) rectangle (9.75,0.5);
    \node at (10.5,7.625) {$\bra*{t_1^{(n)}}$};
    \node at (10.5,4.5) {$\mathbb{M}_{1,2}^{(n)}$};
    \node at (10.5,0) {$\ket*{b_{2}^{(n)}}$};
  \end{tikzpicture},
\end{equation}
where the subscripts refer to fixed-points of different initial states. 
The tensors $A$ and $B$ are given as,
\begin{equation}
  \begin{gathered}
  \begin{tikzpicture}[baseline={([yshift=-0.6ex]current bounding box.center)},scale=0.6]
    \props{0}{0}{white}
    \node at (0,0) {\scalebox{0.6}{$A_{\mathrm{s}}$}};
  \end{tikzpicture}=
  \begin{tikzpicture}[baseline={([yshift=-0.6ex]current bounding box.center)},scale=0.6]
    \props{0}{0}{white}
    \node at (0,0) {\scalebox{0.6}{$A_{\mathrm{F}}$}};
  \end{tikzpicture}=
  \frac{1}{2}
  \begin{tikzpicture}[baseline={([yshift=-0.6ex]current bounding box.center)},scale=0.6]
    \draw[semithick,colLines,rounded corners=3] 
    (-0.5,0.5) -- (-0.25,0.25) -- (-0.25,-0.25) -- (-0.5,-0.5);
    \draw[semithick,colLines,rounded corners=3] 
    (0.5,0.5) -- (0.25,0.25) -- (0.25,-0.25) -- (0.5,-0.5);
  \end{tikzpicture}
  +\frac{1}{2}
  \begin{tikzpicture}[baseline={([yshift=-0.6ex]current bounding box.center)},scale=0.6]
    \draw[semithick,colLines,rounded corners=3] 
    (-0.5,0.5) -- (-0.25,0.25) -- (-0.25,-0.25) -- (-0.5,-0.5);
    \draw[semithick,colLines,rounded corners=3] 
    (0.5,0.5) -- (0.25,0.25) -- (0.25,-0.25) -- (0.5,-0.5);
    \PX{-0.25}{0};
    \PX{0.25}{0};
  \end{tikzpicture},\qquad
  \begin{tikzpicture}[baseline={([yshift=-0.6ex]current bounding box.center)},scale=0.6]
    \gridLinethin{0}{-0.4}{0}{0.4}
    \PX{0}{0};
  \end{tikzpicture}=
  \begin{bmatrix} 0 & 1 \\ 1 & 0
  \end{bmatrix},\qquad
  \begin{tikzpicture}[baseline={([yshift=-0.6ex]current bounding box.center)},scale=0.6]
    \props{0}{0}{white}
    \node at (0,0) {\scalebox{0.6}{$A_{\mathrm{cl}}^{\!(j)}$}};
  \end{tikzpicture}=
  \delta_{b_{j},0}
  \begin{tikzpicture}[baseline={([yshift=-0.6ex]current bounding box.center)},scale=0.6]
    \draw[semithick,colLines,rounded corners=3] 
    (-0.5,0.5) -- (-0.25,0.25) -- (-0.25,-0.25) -- (-0.5,-0.5);
    \draw[semithick,colLines,rounded corners=3] 
    (0.5,0.5) -- (0.25,0.25) -- (0.25,-0.25) -- (0.5,-0.5);
  \end{tikzpicture}
  +\delta_{b_{j},1}
  \begin{tikzpicture}[baseline={([yshift=-0.6ex]current bounding box.center)},scale=0.6]
    \draw[semithick,colLines,rounded corners=3] 
    (-0.5,0.5) -- (-0.25,0.25) -- (-0.25,-0.25) -- (-0.5,-0.5);
    \draw[semithick,colLines,rounded corners=3] 
    (0.5,0.5) -- (0.25,0.25) -- (0.25,-0.25) -- (0.5,-0.5);
    \PX{-0.25}{0};
    \PX{0.25}{0};
  \end{tikzpicture},\\
  \begin{tikzpicture}[baseline={([yshift=-0.6ex]current bounding box.center)},scale=0.6]
    \props{0}{0}{white}
    \node at (0,0) {\scalebox{0.6}{$B_{\mathrm{s}}$}};
  \end{tikzpicture}=
  \begin{tikzpicture}[baseline={([yshift=-0.6ex]current bounding box.center)},scale=0.6]
    \props{0}{0}{white}
    \node at (0,0) {\scalebox{0.6}{$B_{\mathrm{cl}}$}};
  \end{tikzpicture}=
  \begin{tikzpicture}[baseline={([yshift=-0.6ex]current bounding box.center)},scale=0.6]
    \gridLinethin{0.5}{0.5}{-0.5}{-0.5}
    \gridLinethin{0.5}{-0.5}{-0.5}{0.5}
  \end{tikzpicture},\qquad
  \begin{tikzpicture}[baseline={([yshift=-0.6ex]current bounding box.center)},scale=0.6]
    \props{0}{0}{white}
    \node at (0,0) {\scalebox{0.6}{$B_{\mathrm{F}}$}};
  \end{tikzpicture}=
  \begin{tikzpicture}[baseline={([yshift=-0.6ex]current bounding box.center)},scale=0.6]
    \draw[semithick,colLines,rounded corners=3] 
    (-0.5,0.5) -- (-0.25,0.25) -- (-0.25,-0.25) -- (-0.5,-0.5);
    \draw[semithick,colLines,rounded corners=3] 
    (0.5,0.5) -- (0.25,0.25) -- (0.25,-0.25) -- (0.5,-0.5);
  \end{tikzpicture}\,.
  \end{gathered}
\end{equation}
Here s, F, and cl respectively stand for ``solvable", ``flat", and ``classical". Moreover, we note that in the case of the classical configuration $A$ depends on
the precise value of $b_j$ at the time-step $j$.  The initial-state tensors $C$
are given as,
\begin{equation}
  \begin{tikzpicture}[baseline={([yshift=-0.6ex]current bounding box.center)},scale=0.6]
    \propSt{0}{0}{white}
    \node at (0,0) {\scalebox{0.6}{$C_{\mathrm{s}}$}};
  \end{tikzpicture}=
  \frac{1}{2}
  \begin{tikzpicture}[baseline={([yshift=-0.6ex]current bounding box.center)},scale=0.6]
    \draw[semithick,colLines,rounded corners=4] 
    (-0.5,0.5) -- (-0.25,0.25) -- (0.25,0.25) -- (0.5,0.5);
  \end{tikzpicture}+
  \begin{tikzpicture}[baseline={([yshift=-0.6ex]current bounding box.center)},scale=0.6]
    \propSt{0}{0}{white}
    \node at (0,0) {\scalebox{0.6}{$D$}};
  \end{tikzpicture},\qquad
  \begin{tikzpicture}[baseline={([yshift=-0.6ex]current bounding box.center)},scale=0.6]
    \propSt{0}{0}{white}
    \node at (0,0) {\scalebox{0.6}{$C_{\mathrm{cl}}$}};
  \end{tikzpicture}=
  \begin{tikzpicture}[baseline={([yshift=-0.6ex]current bounding box.center)},scale=0.6]
    \gridLinethin{-0.25}{-0.25}{-0.5}{0}
    \node[anchor=center] at (-0.25,-0.5) {\scalebox{0.9}{$s_2$}};
    \gridLinethin{0.25}{-0.25}{0.5}{0}
    \node[anchor=center] at (0.25,-0.5) {\scalebox{0.9}{$\!s_2$}};
  \end{tikzpicture},\qquad
  \begin{tikzpicture}[baseline={([yshift=-0.6ex]current bounding box.center)},scale=0.6]
    \propSt{0}{0}{white}
    \node at (0,0) {\scalebox{0.6}{$C_{\mathrm{F}}$}};
  \end{tikzpicture}=
  \begin{tikzpicture}[baseline={([yshift=-0.6ex]current bounding box.center)},scale=0.6]
    \gridLinethin{-0.25}{-0.25}{-0.5}{0}
    \gridLinethin{0.25}{-0.25}{0.5}{0}
    \MErd{0.25}{-0.25}
    \MEld{-0.25}{-0.25}
  \end{tikzpicture}.
\end{equation}
Here the precise form of $D$ depends on the choice of the solvable initial state, but
it always satisfies
\begin{equation}
  \begin{tikzpicture}[baseline={([yshift=-0.6ex]current bounding box.center)},scale=0.6]
    \draw[colLines,semithick,rounded corners=3] (0,0) -- (-0.75,0.75) -- (0.75,0.75) -- (0,0);
    \begin{scope}
      \clip (-0.33,-0.33) rectangle (0.33,0.33);
      \propSt{0}{0}{white}
    \end{scope}
    \node at (0,0) {\scalebox{0.6}{$D$}};
  \end{tikzpicture}=
  \begin{tikzpicture}[baseline={([yshift=-1.2ex]current bounding box.center)},scale=0.6]
    \draw[colLines,semithick,rounded corners=3] (0,0) -- (-0.75,0.75) -- (0.75,0.75) -- (0,0);
    \begin{scope}
      \clip (-0.33,-0.33) rectangle (0.33,0.33);
      \propSt{0}{0}{white}
    \end{scope}
    \node at (0,0) {\scalebox{0.6}{$D$}};
    \PX{0}{0.75};
  \end{tikzpicture}=0.
\end{equation}
The tensor-network in~\eqref{eq:TNinitialregime} can be easily contracted in
the vertical direction by defining the transfer matrix $\mathbb{M}_{1}^{(n)}$
in the $\mathbb{C}^{\otimes 2n}$ space, and the corresponding boundary vectors
$\bra*{t^{(n)}}$, $\ket*{b_{1,2}^{(n)}}$ (grey boxes
in~\eqref{eq:TNinitialregime}). We note that $\bra*{t^{(n)}}$ is a (left)
eigenvector of $\mathbb{M}_{1,2}^{(n)}$ for any choice of the superscripts $1,2$
(and --- in the case of the classical configuration --- independent of value $b_j$)
\begin{equation}
  \begin{aligned}
    \bra*{t_{\mathrm{s}}^{(n)}}\mathbb{M}_{\mathrm{s},\mathrm{s}}^{(n)}&=
    2^{1-n} \bra*{t_{\mathrm{s}}^{(n)}},&\qquad
    \bra*{t_{\mathrm{s}}^{(n)}}\mathbb{M}_{\mathrm{s},\mathrm{cl}}^{(n)}&=
    2^{1-n} \bra*{t_{\mathrm{s}}^{(n)}},&\qquad
    \bra*{t_{\mathrm{s}}^{(n)}}\mathbb{M}_{\mathrm{s},\mathrm{F}}^{(n)}&=
    \bra*{t_{\mathrm{s}}^{(n)}},\\
    \bra*{t_{\mathrm{cl}}^{(n)}}\mathbb{M}_{\mathrm{cl},\mathrm{s}}^{(n)}&=
    \bra*{t_{\mathrm{cl}}^{(n)}},&
    \bra*{t_{\mathrm{cl}}^{(n)}}\mathbb{M}_{\mathrm{cl},\mathrm{cl}}^{(n)}&=
    \bra*{t_{\mathrm{cl}}^{(n)}},&
    \bra*{t_{\mathrm{cl}}^{(n)}}\mathbb{M}_{\mathrm{cl},\mathrm{F}}^{(n)}&=
    \bra*{t_{\mathrm{cl}}^{(n)}},\\
    \bra*{t_{\mathrm{F}}^{(n)}}\mathbb{M}_{\mathrm{F},\mathrm{s}}^{(n)}&=
    2^{1-n} \bra*{t_{\mathrm{F}}^{(n)}},&
    \bra*{t_{\mathrm{F}}^{(n)}}\mathbb{M}_{\mathrm{F},\mathrm{cl}}^{(n)}&=
    2^{1-n} \bra*{t_{\mathrm{F}}^{(n)}},&
    \bra*{t_{\mathrm{F}}^{(n)}}\mathbb{M}_{\mathrm{F},\mathrm{F}}^{(n)}&=
    \bra*{t_{\mathrm{F}}^{(n)}}.
  \end{aligned}
\end{equation}
To obtain the matrix elements~\eqref{eq:TNinitialregime} the
only thing left to evaluate are the overlaps
$\braket*{t_{1}^{(n)}}{b_{2}^{(n)}}$,
\begin{equation}
  \begin{aligned}
    \braket*{t_{\mathrm{s}}^{(n)}}{b_{\mathrm{s}}^{(n)}}&=2^{1-n},&\qquad
    \braket*{t_{\mathrm{s}}^{(n)}}{b_{\mathrm{cl}}^{(n)}}&=2^{1-n},&\qquad
    \braket*{t_{\mathrm{s}}^{(n)}}{b_{\mathrm{F}}^{(n)}}&=1,\\
    \braket*{t_{\mathrm{cl}}^{(n)}}{b_{\mathrm{s}}^{(n)}}&=1,&\
    \braket*{t_{\mathrm{cl}}^{(n)}}{b_{\mathrm{cl}}^{(n)}}&=1,&
    \braket*{t_{\mathrm{cl}}^{(n)}}{b_{\mathrm{F}}^{(n)}}&=1,\\
    \braket*{t_{\mathrm{F}}^{(n)}}{b_{\mathrm{s}}^{(n)}}&=2^{1-n},&\
    \braket*{t_{\mathrm{F}}^{(n)}}{b_{\mathrm{cl}}^{(n)}}&=2^{1-n},&
    \braket*{t_{\mathrm{F}}^{(n)}}{b_{\mathrm{F}}^{(n)}}&=1,
  \end{aligned}
\end{equation}
which finally give
\begin{equation}
  \begin{aligned}
  \bra{L_{\mathrm{s}}}^{\otimes n} \mathbb{P}_n
    \ket{R_{\mathrm{s}}}^{\otimes n} =
  \bra{L_{\mathrm{s}}}^{\otimes n} \mathbb{P}_n^{\dagger}
    \ket{R_{\mathrm{s}}}^{\otimes n} &= 2^{(1-n)t},\\
  \bra{L_{\mathrm{s}}}^{\otimes n} \mathbb{P}_n
  \ket{R_{\mathrm{F}}}^{\otimes n} =
  \bra{L_{\mathrm{cl}}}^{\otimes n} \mathbb{P}_n^{\dagger}
    \ket{R_{\mathrm{s}}}^{\otimes n}  &= 2^{(1-n)t},\\
  \bra{L_{\mathrm{s}}}^{\otimes n} \mathbb{P}_n
  \ket{R_{\mathrm{cl}}}^{\otimes n} =
  \bra{L_{\mathrm{F}}}^{\otimes n} \mathbb{P}_n^{\dagger}
    \ket{R_{\mathrm{s}}}^{\otimes n} &= 1.
  \end{aligned}
\end{equation}

\subsection{Late-time regime}  \label{sec3}
When $t\to\infty$ we typically expect the finite-subsystem to relax to the
stationary state regardless of the initial state. For special initial states
considered here, the relaxation happens in \emph{finite} time. To see this, we 
consider the following transfer matrix, which gives the exact dynamics of a subsystem
immersed in the large system of solvable states,
\begin{equation}
  \mathcal{C}_x=\frac{1}{4}
  \begin{tikzpicture}[baseline={([yshift=-0.6ex]current bounding box.center)},scale=0.5]
    \foreach\x in {0,2,4,6}{\prop{\x}{1}{FcolU}}
    \foreach\x in {0,2,4}{\prop{1+\x}{0}{FcolU}}
    \fME{-0.5}{1.5}
    \fME{-0.5}{0.5}
    \fME{6.5}{0.5}
    \fME{6.5}{1.5}
    \draw[|<->|] (0,-1) -- (6,-1) node[midway,below] {\scalebox{0.7}{$2x$}};
    
  \end{tikzpicture},
\end{equation}
where the $x$ denotes the size of the subsystem, which in our convention
implies $2x$ open legs (cf.\ the label). For concreteness and simplicity we will
always consider an \emph{even} number of legs, but with minor modifications the
claims can be applied also to odd numbers. In this case the $2x$-th power of
the transfer matrix is expressed as,
\begin{equation}
  \mathcal{C}_x^{2x}=2^{-4x}
  \begin{tikzpicture}[baseline={([yshift=-0.6ex]current bounding box.center)},scale=0.5]
    \foreach \y in {0,2,...,10}{
      \foreach\x in {0,2,4,6}{\prop{\x}{\y+1}{FcolU}}
      \foreach\x in {0,2,4}{\prop{1+\x}{\y}{FcolU}}
      \fME{-0.5}{\y+1.5}
      \fME{-0.5}{\y+0.5}
      \fME{6.5}{\y+1.5}
      \fME{6.5}{\y+0.5}}
    \end{tikzpicture}
    =2^{-2x}
    \begin{tikzpicture}[baseline={([yshift=-0.6ex]current bounding box.center)},scale=0.5]
      \foreach \y in {0,2,...,10}{
        \bendLud{0.5}{\y+0.5}{\y+1.5}
        \Pd{0.3}{\y+1}
        \foreach\x in {2,4,6}{\prop{\x}{\y+1}{FcolU}}
        \foreach\x in {0,2,4}{\prop{1+\x}{\y}{FcolU}}
        \fME{6.5}{\y+0.5}
        \fME{6.5}{\y+1.5}}
      \foreach \y in {1,...,10}{\Pd{1.5}{\y+0.5}}
      \foreach \y in {2,...,9}{\Pd{2.5}{\y+0.5}}
      \foreach \y in {3,...,8}{\Pd{3.5}{\y+0.5}}
      \foreach \y in {4,...,7}{\Pd{4.5}{\y+0.5}}
      \foreach \y in {5,...,6}{\Pd{5.5}{\y+0.5}}
    \end{tikzpicture},
\end{equation}
where $\Pdtext$ is the projector to the subspace with both legs in the same state,
\begin{equation}
\begin{tikzpicture}[baseline={([yshift=-0.6ex]current bounding box.center)},scale=0.5]
    \tgridLine{0}{-0.5}{0}{0.5}
    \Pd{0}{0}
  \end{tikzpicture}=
  \begin{bmatrix}
    1 & 0 & 0 & 0 \\
    0 & 0 & 0 & 0 \\
    0 & 0 & 0 & 0 \\
    0 & 0 & 0 & 1
  \end{bmatrix},
\end{equation}
and the r.h.s.\ follows from
\begin{equation}\label{eq:projPbasicRelations}
  \begin{tikzpicture}[baseline={([yshift=-0.6ex]current bounding box.center)},scale=0.5]
    \prop{0}{0}{FcolU}
    \fME{-0.5}{-0.5}
    \fME{-0.5}{0.5}
  \end{tikzpicture}
  =
  2
  \begin{tikzpicture}[baseline={([yshift=-0.6ex]current bounding box.center)},scale=0.5]
    \bendLud{0.5}{-0.5}{0.5}
    \Pd{0.3}{0}
  \end{tikzpicture},\qquad
  \begin{tikzpicture}[baseline={([yshift=-0.6ex]current bounding box.center)},scale=0.5]
    \tgridLine{0}{0}{-0.75}{0.75}
    \tgridLine{0}{0}{-0.75}{-0.75}
    \prop{0}{0}{FcolU}
    \Pd{-0.5}{-0.5}
    \Pd{-0.5}{0.5}
  \end{tikzpicture}
  =
  \begin{tikzpicture}[baseline={([yshift=-0.6ex]current bounding box.center)},scale=0.5]
    \tgridLine{-0.75}{-0.75}{0.75}{0.75}
    \tgridLine{-0.75}{0.75}{0.75}{-0.75}
    \prop{0}{0}{FcolU}
    \Pd{-0.5}{0.5}
    \Pd{-0.5}{-0.5}
    \Pd{0.5}{0.5}
    \Pd{0.5}{-0.5}
  \end{tikzpicture}.
\end{equation}
Now we note that the following holds,
\begin{equation}\label{eq:projPRERelations}
  \begin{tikzpicture}[baseline={([yshift=-0.6ex]current bounding box.center)},scale=0.5]
    \draw[white] (-0.75,-0.75) -- (0.5,0.5);
    \draw[white] (-0.75,0.75) -- (0.5,-0.5);
    \gridLine{-0.75}{-0.75}{0}{0}
    \prop{0}{0}{FcolU}
    \fME{0.5}{-0.5}
    \fME{0.5}{0.5}
    \Pd{-0.5}{-0.5}
  \end{tikzpicture}=
  \begin{tikzpicture}[baseline={([yshift=-0.6ex]current bounding box.center)},scale=0.5]
    \draw[white] (-0.75,-0.75) -- (0.5,0.5);
    \draw[white] (-0.75,0.75) -- (0.5,-0.5);
    \gridLine{-0.75}{0.75}{0}{0}
    \prop{0}{0}{FcolU}
    \fME{0.5}{-0.5}
    \fME{0.5}{0.5}
    \Pd{-0.5}{0.5}
  \end{tikzpicture}=
  \begin{tikzpicture}[baseline={([yshift=-0.6ex]current bounding box.center)},scale=0.5]
    \gridLine{-0.75}{-0.75}{0}{0}
    \gridLine{-0.75}{0.75}{0}{0}
    \prop{0}{0}{FcolU}
    \fME{0.5}{-0.5}
    \fME{0.5}{0.5}
    \Pd{-0.5}{-0.5}
    \Pd{-0.5}{0.5}
  \end{tikzpicture}=
  \begin{tikzpicture}[baseline={([yshift=-0.6ex]current bounding box.center)},scale=0.5]
    \gridLine{-0.25}{-0.25}{-0.75}{-0.75}
    \gridLine{-0.25}{0.25}{-0.75}{0.75}
    \fME{-0.25}{-0.25}
    \fME{-0.25}{0.25}
  \end{tikzpicture},
\end{equation}
which allows us to simplify the above tensor-network into
\begin{equation}\label{eq:Cxfactorization}
  \mathcal{C}_x^{2x}=
  2^{-2x}
    \begin{tikzpicture}[baseline={([yshift=-0.6ex]current bounding box.center)},scale=0.5]
      \foreach \y in {0,2,...,10}{
        \bendLud{0.5}{\y+0.5}{\y+1.5}
        \Pd{0.3}{\y+1}
        \foreach\x in {2,4}{\prop{\x}{\y+1}{FcolU}}
        \foreach\x in {0,2}{\prop{1+\x}{\y}{FcolU}}
        }
      \foreach \y in {1,...,10}{\Pd{1.5}{\y+0.5}}
      \foreach \y in {2,...,9}{\Pd{2.5}{\y+0.5}}
      \foreach \y in {3,...,8}{\Pd{3.5}{\y+0.5}}
      \foreach \y in {0,2,8,10}{
        \foreach\x in {6}{\prop{\x}{\y+1}{FcolU}}
        \fME{6.5}{\y+1.5}
        \fME{6.5}{\y+0.5}
      }
      \foreach \y in {0,2,4,8,10}{\foreach\x in {4}{\prop{1+\x}{\y}{FcolU}}}
      \fME{5.5}{4.5}
      \fME{4.5}{5.5}
      \fME{4.5}{6.5}
      \fME{5.5}{7.5}
    \end{tikzpicture}=
    2^{-2x}
    \begin{tikzpicture}[baseline={([yshift=-0.6ex]current bounding box.center)},scale=0.5]
      \foreach \y in {0,2,...,10}{
        \bendLud{0.5}{\y+0.5}{\y+1.5}
        \Pd{0.3}{\y+1}}
      \foreach \y in {0,2,4,8,10}{\foreach\x in {0}{\prop{1+\x}{\y}{FcolU}}}
      \foreach \y in {0,2,8,10}{\foreach\x in {2}{\prop{\x}{\y+1}{FcolU}}}
      \foreach \y in {0,2,10}{\foreach\x in {2}{\prop{1+\x}{\y}{FcolU}}}
      \foreach \y in {0,10}{\foreach\x in {4}{\prop{\x}{\y+1}{FcolU}}}
      \foreach \y in {0}{\foreach\x in {4}{\prop{1+\x}{\y}{FcolU}}}
      \gridLine{5.5}{11.5}{5.5}{11}
      \fME{5.5}{11}
      \fME{4.5}{10.5}
      \fME{3.5}{9.5}
      \fME{2.5}{8.5}
      \fME{1.5}{7.5}
      \fME{0.5}{6.5}
      \fME{0.5}{5.5}
      \fME{1.5}{4.5}
      \fME{2.5}{3.5}
      \fME{3.5}{2.5}
      \fME{4.5}{1.5}
      \fME{5.5}{0.5}
    \end{tikzpicture}=
  2^{-2x}\,
    \begin{tikzpicture}[baseline={([yshift=-0.6ex]current bounding box.center)},scale=0.5]
      \foreach \x in {1,...,6}{
        \gridLine{\x}{-0.25}{\x}{-1}
        \gridLine{\x}{0.25}{\x}{1}
        \fME{\x}{-0.25}
        \fME{\x}{0.25}
      }
    \end{tikzpicture},
\end{equation}
where the second equality follows from repeated application
of~\eqref{eq:projPRERelations}, together with the 2nd hierarchy condition
and~\eqref{eq:projPbasicRelations}. To obtain the last equality, we use
unitarity together with the invariance of \fMEtext under \Pdtext.

Eq.~\eqref{eq:Cxfactorization} suggests that application of $C_x$ of at least $2x$ times to \emph{any} initial state results in the maximum entropy state. In particular, this means that starting with the system initialized in the solvable state everywhere except possibly inside a finite subsystem $A$, the density matrix reduced to the subsystem $A$ will be after $2\ell$, $\ell=|A|$, exactly given by the maximum entropy state,
\begin{equation}
  \left.\rho_{A}(t)\right|_{t\ge 2\ell}=\frac{1}{2^{2\ell}}\Id,\qquad \ell=|A|.
\end{equation}

\subsubsection{Late-time regime for the quench from solvable states}

The above statement holds independently of the state \emph{inside} the
subsystem $A$.  However, if the state inside the subsystem $A$  
for a \emph{subset} of solvable initial states, the stationary state is reached
before $2\ell$. In particular, the requirement on the initial state is that it satisfies both the second of Eq.~\eqref{eq:solvablestates} of the main text, and 
\begin{equation} \label{eq:solvableISlatetimeSubset}
  \begin{tikzpicture}[baseline={([yshift=-0.6ex]current bounding box.center)},scale=0.5]
    \gridLine{0}{0}{-0.75}{0.75}
    \prop{0}{0}{FcolU}
    \Pd{-0.5}{0.5}
    \node at (-0.65,-0.65) {\scalebox{0.85}{$\phi_1$}};
    \node at (0.65,-0.65) {\scalebox{0.85}{$\phi_2$}};
  \end{tikzpicture}
  =
  \begin{tikzpicture}[baseline={([yshift=-0.6ex]current bounding box.center)},scale=0.5]
    \gridLine{0}{0}{-0.75}{0.75}
    \gridLine{0}{0}{0.75}{0.75}
    \prop{0}{0}{FcolU}
    \Pd{0.5}{0.5}
    \Pd{-0.5}{0.5}
    \node at (-0.65,-0.65) {\scalebox{0.85}{$\phi_1$}};
    \node at (0.65,-0.65) {\scalebox{0.85}{$\phi_2$}};
  \end{tikzpicture}.
\end{equation}
We start with the diagrammatic representation of the reduced density matrix at time $t=\ell$,
\begin{equation}
  \rho_{A}(\ell)=2^{-2\ell}
  \begin{tikzpicture}[baseline={([yshift=-0.6ex]current bounding box.center)},scale=0.5]
    \foreach \y in {0,2,...,6}{
      \foreach\x in {0,2,...,8}{\prop{\x}{\y+1}{FcolU}}
      \foreach\x in {0,2,...,6}{\prop{1+\x}{\y}{FcolU}}
      \fME{-0.5}{\y+1.5}
      \fME{-0.5}{\y+0.5}
      \fME{8.5}{\y+0.5}
      \fME{8.5}{\y+1.5}}
    \foreach \x in {0,2,...,6}{\node at ({\x+0.35},{-0.65}) {\scalebox{0.85}{$\phi_1$}};}
    \foreach \x in {0,2,...,6}{\node at ({\x+1.65},{-0.65}) {\scalebox{0.85}{$\phi_2$}};}
  \end{tikzpicture}\,.
\end{equation}
Repeatedly using~\eqref{eq:solvableISlatetimeSubset} together with the
manipulations used in the previous subsection, the projectors can be brought
all the way to the right edge, which allows us to start applying the second
hierarchy relation together with the right of~\eqref{eq:solvablestates} of main
text,
\begin{equation}
  \rho_{A}(\ell)= 2^{-\ell}
  \begin{tikzpicture}[baseline={([yshift=-0.6ex]current bounding box.center)},scale=0.5]
    \foreach \y in {0,2,...,6}{
      \foreach\x in {2,4,...,8}{\prop{\x}{\y+1}{FcolU}}
      \foreach\x in {0,2,...,6}{\prop{1+\x}{\y}{FcolU}}
      \bendLud{0.5}{\y+0.5}{\y+1.5}
      \Pd{0.3}{\y+1}
      \fME{8.5}{\y+0.5}
      \fME{8.5}{\y+1.5}}
    \foreach \x in {0,2,...,6}{\node at ({\x+0.35},{-0.65}) {\scalebox{0.85}{$\phi_1$}};}
    \foreach \x in {0,2,...,6}{\node at ({\x+1.65},{-0.65}) {\scalebox{0.85}{$\phi_2$}};}
    \foreach \y in {0,...,6}{\Pd{1.5}{\y+0.5}}
    \foreach \y in {0,...,5}{\Pd{2.5}{\y+0.5}}
    \foreach \y in {0,...,4}{\Pd{3.5}{\y+0.5}}
    \foreach \y in {0,...,3}{\Pd{4.5}{\y+0.5}}
    \foreach \y in {0,...,2}{\Pd{5.5}{\y+0.5}}
    \foreach \y in {0,...,1}{\Pd{6.5}{\y+0.5}}
    \foreach \y in {0}{\Pd{7.5}{\y+0.5}}
  \end{tikzpicture}
  = 2^{-2\ell+1}
  \begin{tikzpicture}[baseline={([yshift=-0.6ex]current bounding box.center)},scale=0.5]
    \foreach \y in {0,2,...,6}{
      \bendLud{0.5}{\y+0.5}{\y+1.5}
    }
    \foreach \y in {0}{\Pd{0.3}{\y+1}}
    \foreach \y in {0,2,4,6}{\prop{1}{\y}{FcolU}}
    \foreach \y in {2,4,6}{\prop{2}{\y+1}{FcolU}}
    \foreach \y in {4,6}{\prop{3}{\y}{FcolU}}
    \foreach \y in {4,6}{\prop{4}{\y+1}{FcolU}}
    \foreach \y in {6}{\prop{5}{\y}{FcolU}}
    \foreach \y in {6}{\prop{6}{\y+1}{FcolU}}
    \fME{1.5}{0.5}
    \fME{1.5}{1.5}
    \fME{2.5}{2.5}
    \fME{3.5}{3.5}
    \fME{4.5}{4.5}
    \fME{5.5}{5.5}
    \fME{6.5}{6.5}
    \gridLine{7.5}{7}{7.5}{7.5}
    \fME{7.5}{7}
    \foreach \x in {0}{\node at ({\x+0.35},{-0.65}) {\scalebox{0.85}{$\phi_1$}};}
    \foreach \x in {0}{\node at ({\x+1.65},{-0.65}) {\scalebox{0.85}{$\phi_2$}};}
  \end{tikzpicture}.
\end{equation}
Now we note that every solution to the second relation of Eq.~\eqref{eq:solvablestates} in the main text satisfies also the following
\begin{equation}
  \begin{tikzpicture}[baseline={([yshift=-0.6ex]current bounding box.center)},scale=0.5]
    \gridLine{0}{0}{-0.75}{0.75}
    \prop{0}{0}{FcolU}
    \Pd{-0.5}{0.5}
    \fME{0.5}{0.5}
    \node at ({-0.65},{-0.65}) {\scalebox{0.85}{$\phi_1$}};
    \node at ({0.65},{-0.65}) {\scalebox{0.85}{$\phi_2$}};
  \end{tikzpicture}=
  \frac{1}{2}
  \begin{tikzpicture}[baseline={([yshift=-0.6ex]current bounding box.center)},scale=0.5]
    \gridLine{0}{0}{0}{0.75}
    \fME{0}{0}
  \end{tikzpicture}.
\end{equation}
This gives us
\begin{equation}
  \rho_{A}(\ell)=2^{-2\ell}
  \begin{tikzpicture}[baseline={([yshift=-0.6ex]current bounding box.center)},scale=0.5]
    \foreach \x in {0,...,7}{
      \gridLine{\x*0.75}{0}{\x*0.75}{0.75}
      \fME{\x*0.75}{0}
    }
  \end{tikzpicture},
\end{equation}
and therefore, in the case of this subset of initial states, the relaxation
happens already at $t=\ell$.

\subsection{Intermediate times}  
\label{sec4}

\subsubsection{Subclass of solvable states}\label{sec:intermediateSolvable}
As the first example we treat the subclass of solvable initial states that
(besides Eq.~\eqref{eq:solvablestates} of the main text) obeys
Eq.~\eqref{eq:solvableISlatetimeSubset}, in the regime where the subsystem size
$\ell=|A|$ is smaller than $2t$, but larger than $t$. In this case, the reduced
density matrix is not yet equal to the stationary one, and at the same time the
powers of the space transfer matrix do not factorize. The reduced density
matrix is in this case given by the following diagram,
\begin{equation}
  \rho_{A}=2^{-2t}
  \begin{tikzpicture}[baseline={([yshift=-0.6ex]current bounding box.center)},scale=0.5]
    \foreach \y in {0,2,...,8}{
      \foreach\x in {0,2,...,12}{\prop{\x}{\y+1}{FcolU}}
      \foreach\x in {0,2,...,10}{\prop{1+\x}{\y}{FcolU}}
      \fME{-0.5}{\y+0.5}
      \fME{-0.5}{\y+1.5}
      \fME{12.5}{\y+0.5}
      \fME{12.5}{\y+1.5}}
    \foreach \x in {1,3,...,11}{
      \node at ({\x-0.65},{-0.7}) {\scalebox{0.85}{$\phi_1$}};
      \node at ({\x+0.65},{-0.7}) {\scalebox{0.85}{$\phi_2$}};
    }
  \end{tikzpicture}\,.
\end{equation}
To express traces of powers of the reduced density matrix we can imagine to
take many copies (replicas) of the diagram above, and connect the top legs in the staggered
way. By introducing dark squares as the notation for $n$ copies of the folded gate (i.e., each
leg now represents $2n$ qubits), and $\fMEtext$, $\fSMEtext$ as symbols for the
two different pairings of the $2n$ legs
\begin{equation}
  \begin{tikzpicture}[baseline={([yshift=-0.6ex]current bounding box.center)},scale=0.5]
    \prop{0}{0}{FncolU}
  \end{tikzpicture}
  =
  \begin{tikzpicture}[baseline={([yshift=-0.6ex]current bounding box.center)},scale=0.5]
    \begin{scope}[shift={(0.15,0.3)}]
      \prop{0}{0}{FcolU}
    \end{scope}
    \begin{scope}[shift={(0.1,0.2)}]
      \prop{0}{0}{FcolU}
    \end{scope}
    \begin{scope}[shift={(0.05,0.1)}]
      \prop{0}{0}{FcolU}
    \end{scope}
    \prop{0}{0}{FcolU}
  \end{tikzpicture},\qquad
  \begin{tikzpicture}[baseline={([yshift=-0.6ex]current bounding box.center)},scale=0.5]
    \gridLine{0}{0}{0}{1}
    \fME{0}{1}
  \end{tikzpicture}
  =
  \begin{tikzpicture}[baseline={([yshift=-0.6ex]current bounding box.center)},scale=0.5]
    \ttopHook{0.3}{1}{0.15}
    \ttopHook{0.9}{1}{0.15}
    \ttopHook{1.5}{1}{0.15}
    \ttopHook{2.1}{1}{0.15}
    \tgridLine{0}{0}{0}{1}
    \tgridLine{0.3}{0}{0.3}{1}
    \tgridLine{0.6}{0}{0.6}{1}
    \tgridLine{0.9}{0}{0.9}{1}
    \tgridLine{1.2}{0}{1.2}{1}
    \tgridLine{1.5}{0}{1.5}{1}
    \tgridLine{1.8}{0}{1.8}{1}
    \tgridLine{2.1}{0}{2.1}{1}
  \end{tikzpicture},\qquad
  \begin{tikzpicture}[baseline={([yshift=-0.6ex]current bounding box.center)},scale=0.5]
    \gridLine{0}{0}{0}{1}
    \fSME{0}{1}
  \end{tikzpicture}=
  \begin{tikzpicture}[baseline={([yshift=-0.6ex]current bounding box.center)},scale=0.5]
    \ttopHook{0.6}{1}{0.15}
    \ttopHook{1.2}{1}{0.15}
    \ttopHook{1.8}{1}{0.15}
    \draw[semithick,colLines,rounded corners=4] (2.1,1) -- (2.1,1.3) -- (0,1.3) -- (0,1);
    \tgridLine{0}{0}{0}{1}
    \tgridLine{0.3}{0}{0.3}{1}
    \tgridLine{0.6}{0}{0.6}{1}
    \tgridLine{0.9}{0}{0.9}{1}
    \tgridLine{1.2}{0}{1.2}{1}
    \tgridLine{1.5}{0}{1.5}{1}
    \tgridLine{1.8}{0}{1.8}{1}
    \tgridLine{2.1}{0}{2.1}{1}
  \end{tikzpicture},
\end{equation}
we can express the trace of $n$-th power of $\rho_{A}(t)$ as
\begin{equation}\label{eq:solvableTracePower}
  \tr[\rho_{A}^n(t)]=2^{-2nt}
  \begin{tikzpicture}[baseline={([yshift=-0.6ex]current bounding box.center)},scale=0.5]
    \foreach \y in {0,2,...,8}{
      \foreach\x in {0,2,...,12}{\prop{\x}{\y+1}{FncolU}}
      \foreach\x in {0,2,...,10}{\prop{1+\x}{\y}{FncolU}}
      \fME{-0.5}{\y+0.5}
      \fME{-0.5}{\y+1.5}
      \fME{12.5}{\y+0.5}
      \fME{12.5}{\y+1.5}}
    \foreach \x in {1,3,...,11}{
      \node at ({\x-0.65},{-0.7}) {\scalebox{0.85}{$\phi_1$}};
      \node at ({\x+0.65},{-0.7}) {\scalebox{0.85}{$\phi_2$}};
      \fSME{\x-0.5}{9.5}
      \fSME{\x+0.5}{9.5}
    }
  \end{tikzpicture}.
\end{equation}
Now we introduce a $n$-generalization of $\Pdtext$ as 
\begin{equation}
  \begin{tikzpicture}[baseline={([yshift=-0.6ex]current bounding box.center)},scale=0.5]
    \gridLine{0}{-0.5}{0}{0.5}
    \Pd{0}{0}
  \end{tikzpicture}=
  \begin{bmatrix}
    1&0&0&0\\
    0&0&0&0\\
    0&0&0&0\\
    0&0&0&1
  \end{bmatrix}^{\otimes n},
\end{equation}
which --- analogously to the $n=1$ case --- satisfies
\begin{equation} \label{eq:relationsProjectorP}
  \begin{tikzpicture}[baseline={([yshift=-0.6ex]current bounding box.center)},scale=0.5]
    \prop{0}{0}{FncolU}
    \fME{-0.5}{-0.5}
    \fME{-0.5}{0.5}
  \end{tikzpicture}=
  2^{n} 
  \begin{tikzpicture}[baseline={([yshift=-0.6ex]current bounding box.center)},scale=0.5]
    \bendLud{0.5}{-0.5}{0.5}
    \Pd{0.3}{0}
  \end{tikzpicture},\qquad
  \begin{tikzpicture}[baseline={([yshift=-0.6ex]current bounding box.center)},scale=0.5]
    \gridLine{0}{0}{-0.75}{-0.75}
    \gridLine{0}{0}{-0.75}{0.75}
    \Pd{-0.5}{-0.5}
    \Pd{-0.5}{0.5}
    \prop{0}{0}{FncolU}
  \end{tikzpicture}=
  \begin{tikzpicture}[baseline={([yshift=-0.6ex]current bounding box.center)},scale=0.5]
    \gridLine{0.75}{0.75}{-0.75}{-0.75}
    \gridLine{0.75}{-0.75}{-0.75}{0.75}
    \Pd{-0.5}{-0.5}
    \Pd{-0.5}{0.5}
    \Pd{0.5}{0.5}
    \Pd{0.5}{-0.5}
    \prop{0}{0}{FncolU}
  \end{tikzpicture}=
  \begin{tikzpicture}[baseline={([yshift=-0.6ex]current bounding box.center)},scale=0.5]
    \draw[white] (-0.75,-0.75) -- (0.75,0.75);
    \gridLine{0}{0}{-0.75}{0.75}
    \gridLine{0}{0}{0.75}{0.75}
    \Pd{-0.5}{0.5}
    \Pd{0.5}{0.5}
    \prop{0}{0}{FncolU}
  \end{tikzpicture}=
  \begin{tikzpicture}[baseline={([yshift=-0.6ex]current bounding box.center)},scale=0.5]
    \draw[white] (-0.75,-0.75) -- (0.75,0.75);
    \gridLine{0}{0}{-0.75}{-0.75}
    \gridLine{0}{0}{0.75}{-0.75}
    \Pd{-0.5}{-0.5}
    \Pd{0.5}{-0.5}
    \prop{0}{0}{FncolU}
  \end{tikzpicture}.
\end{equation}
Repeatedly applying them --- together with the initial-state
condition~\eqref{eq:solvableISlatetimeSubset}, and the unitarity
from the top --- allows us to rewrite~\eqref{eq:solvableTracePower} as
\begin{equation}\label{eq:solvableTracePower2}
  \tr[\rho_{A}^n]=2^{-nt}
  \Big(
  \begin{tikzpicture}[baseline={([yshift=0ex]current bounding box.center)},scale=0.5]
    \gridLine{0}{0}{0}{0.75}
    \fSME{0}{0.75}
    \Pd{0}{0.3}
    \node at ({0},{-0.375}) {\scalebox{0.85}{$\phi_1$}};
  \end{tikzpicture}
  \begin{tikzpicture}[baseline={([yshift=0ex]current bounding box.center)},scale=0.5]
    \gridLine{0}{0}{0}{0.75}
    \fSME{0}{0.75}
    \Pd{0}{0.3}
    \node at ({0},{-0.375}) {\scalebox{0.85}{$\phi_2$}};
  \end{tikzpicture}
  \Big)^{\ell-t}
  \begin{tikzpicture}[baseline={([yshift=-0.6ex]current bounding box.center)},scale=0.5]
    \foreach \x in {3,5,...,9}{
      \gridLine{\x}{0}{\x-0.6}{-0.6}
      \gridLine{\x}{0}{\x+0.6}{-0.6}
    }
    \foreach \y in {0,2,...,8}{
      \foreach\x in {12}{\prop{\x}{\y+1}{FncolU}}
      \foreach\x in {10}{\prop{1+\x}{\y}{FncolU}}
      \fME{12.5}{\y+0.5}
      \fME{12.5}{\y+1.5}}
    \foreach \y in {0,2,4,6}{
      \foreach\x in {10}{\prop{\x}{\y+1}{FncolU}}
      \foreach\x in {8}{\prop{1+\x}{\y}{FncolU}}
    }
    \foreach \y in {0,2,4}{
      \foreach\x in {8}{\prop{\x}{\y+1}{FncolU}}
      \foreach\x in {6}{\prop{1+\x}{\y}{FncolU}}
    }
    \foreach \y in {0,2}{
      \foreach\x in {6}{\prop{\x}{\y+1}{FncolU}}
      \foreach\x in {4}{\prop{1+\x}{\y}{FncolU}}
    }
    \foreach \y in {0}{
      \foreach\x in {4}{\prop{\x}{\y+1}{FncolU}}
      \foreach\x in {2}{\prop{1+\x}{\y}{FncolU}}
    }
    \foreach \x in {3,5,...,11}{
      \node at ({\x-0.6},{-0.8}) {\scalebox{0.85}{$\phi_1$}};
      \node at ({\x+0.8},{-0.8}) {\scalebox{0.85}{$\phi_2$}};
    }
    \foreach \x in {0,...,9}{\fSME{11.5-\x}{9.5-\x}}
    \foreach \x in {2,4,6,8}{\Pd{\x+0.6}{-0.4}}
    \foreach \x in {3,5,7,9}{\Pd{\x+0.4}{-0.4}}
    \draw [|<->|] (2,-1.5) -- (10,-1.5) node[midway,below] {\scalebox{0.8}{$4t-2\ell$}};
    \draw [|<->|] (10,-1.5) -- (12,-1.5) node[midway,below] {\scalebox{0.8}{$2\ell-2t$}};
  \end{tikzpicture}\, .
\end{equation}
To progress from here, we note that the condition~\eqref{eq:solvableISlatetimeSubset} is satisfied whenever at least one of $\phi_1$ and $\phi_2$ is a classical state, i.e., it is either $[1,0]$ or $[0,1]$. Using now 
\begin{equation}
  \begin{tikzpicture}[baseline={([yshift=0ex]current bounding box.center)},scale=0.5]
    \prop{0}{0}{FcolU}
    \node at (-0.75,-0.75) {\scalebox{0.7}{$\underline{s}$}};
    \fME{-0.5}{0.5}
  \end{tikzpicture}
  = \frac{1}{2}
  \begin{tikzpicture}[baseline={([yshift=-0.6ex]current bounding box.center)},scale=0.5]
    \prop{0}{0}{FcolU}
    \fME{-0.5}{-0.5}
    \fME{-0.5}{0.5}
  \end{tikzpicture},
  \qquad
  \begin{tikzpicture}[baseline={([yshift=0ex]current bounding box.center)},scale=0.5]
    \prop{0}{0}{FcolU}
    \fME{-0.5}{0.5}
    \node at (0.75,-0.75) {\scalebox{0.7}{$\underline{s}$}};
    \node at (-0.75,-0.75) {\scalebox{0.7}{$\phi$}};
  \end{tikzpicture}
  =
  \frac{1}{2}   \begin{tikzpicture}[baseline={([yshift=0ex]current bounding box.center)},scale=0.5]
    \gridLine{0}{0}{0}{0.75}
    \fSME{0}{0.75}
    \Pd{0}{0.25}
    \node at (0,-0.25) {\scalebox{0.7}{$\phi$}};
  \end{tikzpicture}\begin{tikzpicture}[baseline={([yshift=0ex]current bounding box.center)},scale=0.5]
    \prop{0}{0}{FcolU}
    \fME{-0.5}{-0.5}
    \fME{-0.5}{0.5}
    \node at (0.75,-0.75) {\scalebox{0.7}{$\underline{s}$}};
  \end{tikzpicture},
\end{equation}
we find the following generalisation of Eq.~\eqref{eq:solvablestates} of the main text 
\begin{equation}\label{eq:genclassrel}
  \begin{tikzpicture}[baseline={([yshift=-0.6ex]current bounding box.center)},scale=0.5]
    \prop{0}{0}{FncolU}
    \prop{1}{1}{FncolU}
    \fSME{-0.5}{0.5}
    \fSME{0.5}{1.5}
    \node at (-0.75,-0.75) {\scalebox{0.7}{$\underline{s}$}};
  \end{tikzpicture}=
  2^{-n}
  \begin{tikzpicture}[baseline={([yshift=-0.6ex]current bounding box.center)},scale=0.5]
    \prop{0}{0}{FncolU}
    \prop{1}{1}{FncolU}
    \fSME{-0.5}{0.5}
    \fSME{0.5}{1.5}
    \fSME{-0.5}{-0.5}
  \end{tikzpicture}=
  2^{-n}
  \begin{tikzpicture}[baseline={([yshift=-0.6ex]current bounding box.center)},scale=0.5]
    \prop{1}{1}{FncolU}
    \gridLine{0.5}{-0.25}{1.5}{-0.25}
    \fSME{0.5}{0.5}
    \fSME{0.5}{1.5}
    \fSME{0.5}{-0.25}
  \end{tikzpicture},\qquad
  \begin{tikzpicture}[baseline={([yshift=-0.6ex]current bounding box.center)},scale=0.5]
    \prop{0}{0}{FncolU}
    \prop{1}{1}{FncolU}
    \fSME{-0.5}{0.5}
    \fSME{0.5}{1.5}
    \node at (-0.75,-0.75) {\scalebox{0.7}{$\phi$}};
    \node at (0.75,-0.75) {\scalebox{0.7}{$\underline{s}$}};
  \end{tikzpicture}=
  \begin{tikzpicture}[baseline={([yshift=0ex]current bounding box.center)},scale=0.5]
    \gridLine{0}{0}{0}{0.75}
    \fSME{0}{0.75}
    \node at (0,-0.25) {\scalebox{0.7}{$\phi$}};
  \end{tikzpicture}
  2^{-n}
  \begin{tikzpicture}[baseline={([yshift=-0.6ex]current bounding box.center)},scale=0.5]
    \prop{0}{0}{FncolU}
    \prop{1}{1}{FncolU}
    \fSME{-0.5}{0.5}
    \fSME{0.5}{1.5}
    \fSME{-0.5}{-0.5}
    \node at (0.75,-0.75) {\scalebox{0.7}{$\underline{s}$}};
  \end{tikzpicture}.
\end{equation}
Here and in the following we denote by $\underline{s}$ a classical configuration on all the $2n$ copies. These equations in particular imply
\begin{equation}
  \begin{tikzpicture}[baseline={([yshift=-0.6ex]current bounding box.center)},scale=0.5]
    \gridLine{0}{0}{-0.6}{-0.6}
    \gridLine{0}{0}{0.6}{-0.6}
    \prop{0}{0}{FncolU}
    \prop{1}{1}{FncolU}
    \fSME{-0.5}{0.5}
    \fSME{0.5}{1.5}
    \Pd{-0.4}{-0.4}
    \Pd{0.4}{-0.4}
    \node at (-0.75,-0.75) {\scalebox{0.7}{$\phi_1$}};
    \node at (0.75,-0.75) {\scalebox{0.7}{$\phi_2$}};
  \end{tikzpicture}=
  2^{-n}
  \begin{tikzpicture}[baseline={([yshift=0ex]current bounding box.center)},scale=0.5]
    \gridLine{0}{0}{0}{0.75}
    \fSME{0}{0.75}
    \Pd{0}{0.25}
    \node at (0,-0.25) {\scalebox{0.7}{$\phi_1$}};
  \end{tikzpicture}
  \begin{tikzpicture}[baseline={([yshift=0ex]current bounding box.center)},scale=0.5]
    \gridLine{0}{0}{0}{0.75}
    \fSME{0}{0.75}
    \Pd{0}{0.25}
    \node at (0,-0.25) {\scalebox{0.7}{$\phi_2$}};
  \end{tikzpicture}
  \begin{tikzpicture}[baseline={([yshift=-0.6ex]current bounding box.center)},scale=0.5]
    \prop{0}{0}{FncolU}
    \fSME{-0.5}{0.5}
    \fSME{-0.5}{-0.5}
  \end{tikzpicture}.
\end{equation}
This reduces the triangle in~\eqref{eq:solvableTracePower2} into
\begin{equation}
  \tr[\rho_{A}^n(t)]=
  2^{-2tn}
  \left(
  \begin{tikzpicture}[baseline={([yshift=0ex]current bounding box.center)},scale=0.5]
    \gridLine{0}{0}{0}{0.75}
    \fSME{0}{0.75}
    \Pd{0}{0.25}
    \node at (0,-0.25) {\scalebox{0.7}{$\phi_1$}};
  \end{tikzpicture}
  \begin{tikzpicture}[baseline={([yshift=0ex]current bounding box.center)},scale=0.5]
    \gridLine{0}{0}{0}{0.75}
    \fSME{0}{0.75}
    \Pd{0}{0.25}
    \node at (0,-0.25) {\scalebox{0.7}{$\phi_2$}};
  \end{tikzpicture}\right)^t
  \left(
  \begin{tikzpicture}[baseline={([yshift=-0.6ex]current bounding box.center)},scale=0.5]
    \foreach\x in {12}{\prop{\x}{0}{FncolU}}
    \foreach \y in {0}{
      \fME{12.5}{\y-0.5}
      \fME{12.5}{\y+0.5}
      \fSME{11.5}{\y+0.5}
      \fSME{11.5}{\y-0.5}
    }
  \end{tikzpicture}
  \right)^{t}.
\end{equation}
The only thing left to do is to combine all the relevant factors
\begin{equation}
  \begin{tikzpicture}[baseline={([yshift=-0.6ex]current bounding box.center)},scale=0.5]
    \foreach\x in {12}{\prop{\x}{0}{FncolU}}
    \foreach \y in {0}{
      \fME{12.5}{\y-0.5}
      \fME{12.5}{\y+0.5}
      \fSME{11.5}{\y+0.5}
      \fSME{11.5}{\y-0.5}
    }
  \end{tikzpicture}= 2^{n+1},\qquad
  \begin{tikzpicture}[baseline={([yshift=0ex]current bounding box.center)},scale=0.5]
    \gridLine{0}{0}{0}{0.75}
    \fSME{0}{0.75}
    \Pd{0}{0.25}
    \node at (0,-0.25) {\scalebox{0.7}{$\phi_1$}};
  \end{tikzpicture}
  \begin{tikzpicture}[baseline={([yshift=0ex]current bounding box.center)},scale=0.5]
    \gridLine{0}{0}{0}{0.75}
    \fSME{0}{0.75}
    \Pd{0}{0.25}
    \node at (0,-0.25) {\scalebox{0.7}{$\phi_2$}};
  \end{tikzpicture}=2^{-n+1},
\end{equation}
where we remark that the second equality holds for all initial states that are both
solvable and satisfy~\eqref{eq:solvableISlatetimeSubset} (i.e., they are in the subset considered here).
This finally gives
\begin{equation}
  \tr[\rho_{A}^{n}(t)]=2^{2t(1-n)}.
\end{equation}
As a finaly point, we remark that here we never explicitly used translational
invariance of the initial state, and pairs $(\phi_1,\phi_2)$ could be chosen
independently for all pairs of sites (as long as the relevant conditions are
satisfied) without changing the result.

\subsubsection{Classical subregion in a system prepared in a solvable state}
Let us now consider entanglement entropy between the subsystem $A$ and the
rest, where the rest is prepared in the solvable state, while $A$ is
initialized in a classical configuration. First we treat the regime $t<\ell<2t$,
and we denote the classical configuration by $[s_1,s_2,\ldots,s_{2\ell}]$.
Diagrammatically, the trace of powers of the reduced density matrix takes the
following form
\begin{equation}
  \tr[\rho_{A}^n(t)]=2^{-2nt}
  \begin{tikzpicture}[baseline={([yshift=-0.6ex]current bounding box.center)},scale=0.5]
    \foreach \y in {0,2,...,8}{
      \foreach\x in {0,2,...,12}{\prop{\x}{\y+1}{FncolU}}
      \foreach\x in {0,2,...,10}{\prop{1+\x}{\y}{FncolU}}
      \fME{-0.5}{\y+0.5}
      \fME{-0.5}{\y+1.5}
      \fME{12.5}{\y+0.5}
      \fME{12.5}{\y+1.5}}
    \node at (0.5,-0.875) {\scalebox{0.7}{$\underline{s_1}$}};
    \node at (1.5,-0.875) {\scalebox{0.7}{$\underline{s_2}$}};
    \node at (2.5,-0.875) {\scalebox{0.7}{$\underline{s_3}$}};
    \node at (3.5,-0.875) {\scalebox{1}{$\cdots$}};
    \node at (4.5,-0.875) {\scalebox{1}{$\cdots$}};
    \node at (5.5,-0.875) {\scalebox{1}{$\cdots$}};
    \node at (6.5,-0.875) {\scalebox{1}{$\cdots$}};
    \node at (7.5,-0.875) {\scalebox{1}{$\cdots$}};
    \node at (8.5,-0.875) {\scalebox{1}{$\cdots$}};
    \node at (9.5,-0.875) {\scalebox{1}{$\cdots$}};
    \node at (10.5,-0.875) {\scalebox{0.7}{$\underline{s_{2\ell-1}}$}};
    \node at (11.5,-0.875) {\scalebox{0.7}{$\underline{s_{2\ell}}$}};
    \foreach \x in {1,3,...,11}{
      \fSME{\x-0.5}{9.5}
      \fSME{\x+0.5}{9.5}
    }
  \end{tikzpicture}.
\end{equation}
Using the same simplifications as in the previous section we obtain
\begin{equation}
  \tr[\rho_{A}^n(t)]=
  2^{-2tn}
  \left(
  \begin{tikzpicture}[baseline={([yshift=-0.6ex]current bounding box.center)},scale=0.5]
    \foreach\x in {12}{\prop{\x}{0}{FncolU}}
    \foreach \y in {0}{
      \fME{12.5}{\y-0.5}
      \fME{12.5}{\y+0.5}
      \fSME{11.5}{\y+0.5}
      \fSME{11.5}{\y-0.5}
    }
  \end{tikzpicture}
  \right)^{t}=2^{t(1-n)}.
\end{equation}

We now consider the regime $\frac{t}{2}<\ell<t$ and compute the following 
\begin{equation}
  \tr[\rho_{A}^n(t)]=2^{-2nt}
  \begin{tikzpicture}[baseline={([yshift=-0.6ex]current bounding box.center)},scale=0.5]
    \foreach \y in {0,2,...,8,10}{
      \foreach\x in {0,2,...,8}{\prop{\x}{\y+1}{FncolU}}
      \foreach\x in {0,2,...,6}{\prop{1+\x}{\y}{FncolU}}
      \fME{-0.5}{\y+0.5}
      \fME{-0.5}{\y+1.5}
      \fME{8.5}{\y+0.5}
      \fME{8.5}{\y+1.5}
    }
    \node at (0.5,-0.875) {\scalebox{0.7}{$\underline{s_1}$}};
    \node at (1.5,-0.875) {\scalebox{0.7}{$\underline{s_2}$}};
    \node at (2.5,-0.875) {\scalebox{0.7}{$\underline{s_3}$}};
    \node at (3.5,-0.875) {\scalebox{1}{$\cdots$}};
    \node at (4.5,-0.875) {\scalebox{1}{$\cdots$}};
    \node at (5.5,-0.875) {\scalebox{1}{$\cdots$}};
    \node at (6.5,-0.875) {\scalebox{0.7}{$\underline{s_{2\ell-1}}$}};
    \node at (7.5,-0.875) {\scalebox{0.7}{$\underline{s_{2\ell}}$}};
    \foreach \x in {1,3,...,7}{
      \fSME{\x-0.5}{11.5}
      \fSME{\x+0.5}{11.5}
    }
  \end{tikzpicture}
  =2^{-nt}
  \begin{tikzpicture}[baseline={([yshift=-0.6ex]current bounding box.center)},scale=0.5]

    \foreach\x in {8}{\prop{\x}{0}{FncolU}}
    \foreach\x in {6}{\prop{1+\x}{0+1}{FncolU}}
    \foreach\x in {6,8}{\prop{\x}{2}{FncolU}}
    \foreach\x in {4,6}{\prop{1+\x}{2+1}{FncolU}}
    \foreach\x in {4,6,8}{\prop{\x}{4}{FncolU}}
    \foreach\x in {2,4,6}{\prop{1+\x}{4+1}{FncolU}}
    \foreach\x in {2,4,6,8}{\prop{\x}{6}{FncolU}}
    \foreach\x in {0,2,4,6}{\prop{1+\x}{6+1}{FncolU}}

    \foreach \y in {8,10}{
      \bendLud{0.5}{\y-0.5}{\y+0.5}
      \Pd{0.3}{\y}
    }

    \foreach\y in {0,2,...,6}{
      \fME{8.5}{\y+0.5}
      \fME{8.5}{\y-0.5}
    }

    \foreach \y in {8}{
      \foreach\x in {2,6,8}{\prop{\x}{\y}{FncolU}}
      \foreach\x in {0,6}{\prop{1+\x}{\y+1}{FncolU}}
      \fME{8.5}{\y+0.5}
      \fME{8.5}{\y-0.5}}
    \foreach \y in {10}{
      \foreach\x in {8}{\prop{\x}{\y}{FncolU}}
      \fME{8.5}{\y+0.5}
      \fME{8.5}{\y-0.5}}
    \fSME{1-0.5}{10.5}
    \fSME{7+0.5}{10.5}
    \fSME{1+0.5}{9.5}
    \fSME{7-0.5}{9.5}
    \fSME{2+0.5}{8.5}
    \fSME{6-0.5}{8.5}
    \fSME{3+0.5}{7.5}
    \fSME{5-0.5}{7.5}

    \node at (0.5,6.125) {\scalebox{0.7}{$\underline{s^\prime_1}$}};
    \node at (1.5,5.125) {\scalebox{1}{$\cdots$}};
    \node at (2.5,4.125) {\scalebox{0.7}{$\underline{s^\prime_{2\ell-t+1}}$}};
    \node at (3.5,3.125) {\scalebox{1}{$\cdots$}};
    \node at (4.5,2.125) {\scalebox{1}{$\cdots$}};
    \node at (5.5,1.125) {\scalebox{1}{$\cdots$}};
    \node at (6.5,0.125) {\scalebox{0.7}{$\underline{s^\prime_{2\ell-1}}$}};
    \node at (7.5,-0.875) {\scalebox{0.7}{$\underline{s_{2\ell}}$}};

  \end{tikzpicture}\;.
  \label{eq:thirdreg1}
\end{equation}
The r.h.s.\ follows by using unitarity together with the first
of~\eqref{eq:relationsProjectorP}, and then using the deterministic rule to
evolve the initial classical state $[s_j]_j$ to the classical state
$[s^{\prime}_j]_j$ on the diagonal. We can now first remove the projectors on the left by continuously applying the r.h.s.\ of~\eqref{eq:relationsProjectorP}, and then by combining unitarity and Eq.~\eqref{eq:solvablestates} the diagram contracts fully: 
\begin{equation}
  \tr[\rho_{A}^n(t)]
  = 2^{-nt}
  \begin{tikzpicture}[baseline={([yshift=-0.6ex]current bounding box.center)},scale=0.5]

    \foreach\x in {8}{\prop{\x}{0}{FncolU}}
    \foreach\x in {6}{\prop{1+\x}{0+1}{FncolU}}
    \foreach\x in {6,8}{\prop{\x}{2}{FncolU}}
    \foreach\x in {4,6}{\prop{1+\x}{2+1}{FncolU}}
    \foreach\x in {4,6,8}{\prop{\x}{4}{FncolU}}
    \foreach\x in {2,4,6}{\prop{1+\x}{4+1}{FncolU}}
    \foreach\x in {2,4,6,8}{\prop{\x}{6}{FncolU}}
    \foreach\x in {0,2,4,6}{\prop{1+\x}{6+1}{FncolU}}

    \foreach \y in {8}{\bendLud{0.5}{\y-0.5}{\y+0.5}}
    \foreach\y in {0,2,...,6}{
      \fME{8.5}{\y+0.5}
      \fME{8.5}{\y-0.5}
    }

    \foreach \y in {8}{
      \foreach\x in {2,6,8}{\prop{\x}{\y}{FncolU}}
      \foreach\x in {0,6}{\prop{1+\x}{\y+1}{FncolU}}
      \fME{8.5}{\y+0.5}
      \fME{8.5}{\y-0.5}}
    \foreach \y in {10}{
      \foreach\x in {8}{\prop{\x}{\y}{FncolU}}
      \fME{8.5}{\y+0.5}
      \fME{8.5}{\y-0.5}}
    \fSME{1-0.5}{9.5}
    \fSME{7+0.5}{10.5}
    \fSME{1+0.5}{9.5}
    \fSME{7-0.5}{9.5}
    \fSME{2+0.5}{8.5}
    \fSME{6-0.5}{8.5}
    \fSME{3+0.5}{7.5}
    \fSME{5-0.5}{7.5}

    \node at (0.5,6.125) {\scalebox{0.7}{$\underline{s^\prime_1}$}};
    \node at (1.5,5.125) {\scalebox{1}{$\cdots$}};
    \node at (2.5,4.125) {\scalebox{0.7}{$\underline{s^\prime_{2l-t+1}}$}};
    \node at (3.5,3.125) {\scalebox{1}{$\cdots$}};
    \node at (4.5,2.125) {\scalebox{1}{$\cdots$}};
    \node at (5.5,1.125) {\scalebox{1}{$\cdots$}};
    \node at (6.5,0.125) {\scalebox{0.7}{$\underline{s^\prime_{2\ell-1}}$}};
    \node at (7.5,-0.875) {\scalebox{0.7}{$\underline{s_{2\ell}}$}};
  \end{tikzpicture}
  = 2^{-nt}
  \begin{tikzpicture}[baseline={([yshift=-0.6ex]current bounding box.center)},scale=0.5]

    \foreach\x in {8}{\prop{\x}{0}{FncolU}}
    \foreach\x in {6}{\prop{1+\x}{0+1}{FncolU}}
    \foreach\x in {6,8}{\prop{\x}{2}{FncolU}}
    \foreach\x in {4,6}{\prop{1+\x}{2+1}{FncolU}}
    \foreach\x in {4,6,8}{\prop{\x}{4}{FncolU}}
    \foreach\x in {2,4,6}{\prop{1+\x}{4+1}{FncolU}}
    \foreach\x in {4,6,8}{\prop{\x}{6}{FncolU}}
    \foreach\x in {4,6}{\prop{1+\x}{6+1}{FncolU}}

    \foreach \y in {8}{
      \foreach\x in {6,8}{\prop{\x}{\y}{FncolU}}
      \foreach\x in {6}{\prop{1+\x}{\y+1}{FncolU}}
      \fME{8.5}{\y+0.5}
      \fME{8.5}{\y-0.5}}
    \foreach \y in {10}{
      \foreach\x in {8}{\prop{\x}{\y}{FncolU}}
      \fME{8.5}{\y+0.5}
      \fME{8.5}{\y-0.5}}

    \foreach\y in {0,2,...,6}{
      \fME{8.5}{\y+0.5}
      \fME{8.5}{\y-0.5}
    }

    \fSME{7+0.5}{10.5}
    \fSME{7-0.5}{9.5}
    \fSME{6-0.5}{8.5}
    \fSME{5-0.5}{7.5}
    \fSME{4-0.65}{6+0.65}
    \fSME{3-0.65}{5 +0.65}
    \node at (2.5,4.125) {\scalebox{0.7}{$\underline{s^\prime_{2l-t+1}}$}};
    \node at (3.5,3.125) {\scalebox{1}{$\cdots$}};
    \node at (4.5,2.125) {\scalebox{1}{$\cdots$}};
    \node at (5.5,1.125) {\scalebox{1}{$\cdots$}};
    \node at (6.5,0.125) {\scalebox{0.7}{$\underline{s^\prime_{2\ell-1}}$}};
    \node at (7.5,-0.875) {\scalebox{0.7}{$\underline{s_{2\ell}}$}};
    \draw[|<->|] (1.75,-1.5) -- (7.75,-1.5) node[midway,below] {\scalebox{0.9}{$t-1$}};
  \end{tikzpicture}\, 
  =
  2^{-2nt+n}
  \left(
  \begin{tikzpicture}[baseline={([yshift=-0.6ex]current bounding box.center)},scale=0.5]
    \prop{0}{0}{FncolU}
    \fSME{-0.5}{-0.5}
    \fSME{-0.5}{0.5}
    \fME{0.5}{-0.5}
    \fME{0.5}{0.5}
  \end{tikzpicture}
  \right)^{t-1}
  \begin{tikzpicture}[baseline={([yshift=0ex]current bounding box.center)},scale=0.5]
    \prop{0}{0}{FncolU}
    \fSME{-0.5}{0.5}
    \fME{0.5}{-0.5}
    \fME{0.5}{0.5}
    \node at (-0.5,-0.875) {\scalebox{0.7}{$\underline{s_{2\ell}}$}};
  \end{tikzpicture}.
\end{equation}
After noting
\begin{equation}
  \begin{tikzpicture}[baseline={([yshift=0ex]current bounding box.center)},scale=0.5]
    \prop{0}{0}{FncolU}
    \fSME{-0.5}{0.5}
    \fME{0.5}{-0.5}
    \fME{0.5}{0.5}
    \node at (-0.5,-0.875) {\scalebox{0.7}{$\underline{s_{2\ell}}$}};
  \end{tikzpicture}=2,
\end{equation}
we can finally gather all the factors together and obtain
\begin{equation}
  \tr[\rho_{A}^n(t)] = 2^{t(1-n)} \,. 
\end{equation}

\subsubsection{Region of flat states in a system prepared in the solvable state}

Now we consider a quench from the system prepared in the solvable state everywhere, except
for a finite region $A$ which is initialized in the flat state, and we are interested in
the entanglement entropy between $A$ and the rest at some later time $t$. Since the flat
state is locally invariant under time-evolution, the transfer matrix factorizes into fixed
points for all times $\ell\le|A|$ (and not only for $t\le \ell/2$). What remains to be understood
is the intermediate regime with $\ell<t<2\ell$. The trace of the $n$-th power of the reduced density
matrix is given as
\begin{equation}
  \tr[\rho_{A}^n]
  =2^{-2nt}
  \begin{tikzpicture}[baseline={([yshift=-0.6ex]current bounding box.center)},scale=0.5]
    \foreach \y in {0,2,...,10}{
      \foreach\x in {0,2,...,8}{\prop{\x}{\y+1}{FncolU}}
      \foreach\x in {0,2,...,6}{\prop{1+\x}{\y}{FncolU}}
      \fME{-0.5}{\y+0.5}
      \fME{-0.5}{\y+1.5}
      \fME{8.5}{\y+0.5}
      \fME{8.5}{\y+1.5}}
    \foreach \x in {0,2,...,6}{\MErd{\x+0.5}{-0.5}}
    \foreach \x in {0,2,...,6}{\MEld{\x+1.5}{-0.5}}
    \foreach \x in {1,3,...,7}{
      \fSME{\x-0.5}{11.5}
      \fSME{\x+0.5}{11.5}
    }
  \end{tikzpicture}
  = 2^{n (\ell-2t)}
  \begin{tikzpicture}[baseline={([yshift=-0.6ex]current bounding box.center)},scale=0.5]
    \foreach\x in {0,8}{\prop{\x}{10}{FncolU}}
    \foreach\x in {0,6}{\prop{1+\x}{9}{FncolU}}
    \foreach\x in {0,2,6,8}{\prop{\x}{8}{FncolU}}
    \foreach\x in {0}{\prop{\x}{0}{FncolU}}
    \foreach\x in {0}{\prop{\x+1}{1}{FncolU}}
    \foreach\x in {0,2}{\prop{\x}{2}{FncolU}}
    \foreach\x in {0,2}{\prop{\x+1}{3}{FncolU}}
    \foreach\x in {0,2,4}{\prop{\x}{4}{FncolU}}
    \foreach\x in {0,2,4}{\prop{\x+1}{5}{FncolU}}
    \foreach\x in {0,2,4,6}{\prop{\x}{6}{FncolU}}
    \foreach\x in {0,2,4,6}{\prop{\x+1}{7}{FncolU}}
    \foreach \y in {7,...,10}{\fME{8.5}{\y+0.5}}
    \foreach \y in {0,2,...,10}{
      \fME{-0.5}{\y-0.5}
      \fME{-0.5}{\y+0.5}
    }
    \foreach \x in {0,...,7}{\MEld{0.5+\x}{-0.5+\x}}
    \foreach \x in {1,...,4}{
      \fSME{\x-0.5}{11.5-\x}
      \fSME{9-\x-0.5}{11.5-\x}
    }
    \draw[|<->|] (9,7) -- (9,11) node[midway,right] {\scalebox{0.8}{$2t-2\ell$}};
  \end{tikzpicture},
\end{equation}
where the r.h.s.\ follows by applying unitarity from the top and the local invariance
of the flat state. Using now relations~\eqref{eq:relationsProjectorP}, we introduce
projectors $\Pdtext$ on the left edge and bring them down to the diagonal. Then, using the unitarity of gates from top left, we end up with a diagonal strip
\begin{equation} \label{eq:FlatStateIntermediate2}
  \tr[\rho_{A}^n(t)]
  =2^{n(\ell-t)}
  \begin{tikzpicture}[baseline={([yshift=-0.6ex]current bounding box.center)},scale=0.5]
    \foreach\x in {8}{\prop{\x}{10}{FncolU}}
    \foreach\x in {0,6}{\prop{1+\x}{9}{FncolU}}
    \foreach\x in {2,6,8}{\prop{\x}{8}{FncolU}}
    \foreach\x in {0}{\prop{\x+1}{1}{FncolU}}
    \foreach\x in {2}{\prop{\x}{2}{FncolU}}
    \foreach\x in {0,2}{\prop{\x+1}{3}{FncolU}}
    \foreach\x in {2,4}{\prop{\x}{4}{FncolU}}
    \foreach\x in {0,2,4}{\prop{\x+1}{5}{FncolU}}
    \foreach\x in {2,4,6}{\prop{\x}{6}{FncolU}}
    \foreach\x in {0,2,4,6}{\prop{\x+1}{7}{FncolU}}
    \foreach \y in {7,...,10}{\fME{8.5}{\y+0.5}}
    \foreach \y in {2,4,6,8}{\bendLud{0.5}{\y-0.5}{\y+0.5}}
    \gridLine{0.5}{0.5}{0.25}{0.25}
    \Pd{0.5}{0.5}
    \MErd{0.25}{0.25}
    \foreach \x in {1,...,5}{
      \gridLine{0.5+\x}{-0.5+\x}{0.75+\x}{-0.75+\x}
      \Pd{0.5+\x}{-0.5+\x}
      \MEld{0.75+\x}{-0.75+\x}
    }
    \foreach \x in {6,7}{\MEld{0.5+\x}{-0.5+\x}}
    \foreach \x in {1,...,4}{\fSME{9-\x-0.5}{11.5-\x}}
    \foreach \x in {2,...,4}{\fSME{\x-0.5}{11.5-\x}}
    \fSME{0.5}{9.5}
    \draw[|<->|] (0,-0.5) -- (5.99,-0.5) node[midway,below] {\scalebox{0.9}{$t$}};
    \draw[|<->|] (6.01,-0.5) -- (7.5,-0.5) node[midway,below] {\scalebox{0.9}{$2\ell-t$}};
    \draw[|<->|] (0,11) -- (3.95,11) node[midway,above] {\scalebox{0.9}{$\ell$}};
    \draw[|<->|] (4.05,11) -- (8,11) node[midway,above] {\scalebox{0.9}{$\ell$}};
  \end{tikzpicture}
  =
  \underbrace{2^{n(\ell-t)}
  \left(
  \begin{tikzpicture}[baseline={([yshift=-0.6ex]current bounding box.center)},scale=0.5]
    \gridLine{0}{0.25}{0}{1}
    \Pd{0}{0.5}
    \MEh{0}{0.25}
    \fSME{0}{1}
  \end{tikzpicture}
  \right)^{\ell}}_{2^{\ell-nt}}
  \begin{tikzpicture}[baseline={([yshift=-0.6ex]current bounding box.center)},scale=0.5]
    \gridLine{0.5}{-0.5}{0.25}{-0.75}
    \gridLine{-0.5}{0.5}{-0.75}{0.25}
    \prop{0}{1}{FncolU}
    \prop{1}{2}{FncolU}
    \prop{2}{3}{FncolU}
    \prop{3}{4}{FncolU}
    \prop{1}{0}{FncolU}
    \prop{2}{1}{FncolU}
    \prop{3}{2}{FncolU}
    \fSME{-0.5}{1.5}
    \fSME{0.5}{2.5}
    \fSME{1.5}{3.5}
    \fSME{2.5}{4.5}
    \fME{3.5}{4.5}
    \fME{3.5}{3.5}
    \fME{3.5}{2.5}
    \fME{3.5}{1.5}
    \MEld{2.5}{0.5}
    \MEld{1.5}{-0.5}
    \Pd{0.5}{-0.5}
    \Pd{-0.5}{0.5}
    \MErd{0.25}{-0.75}
    \MErd{-0.75}{0.25}
    \draw[|<->|] (-1,5) -- (3,5) node[midway,above] {\scalebox{0.9}{$\ell$}};
    \draw[|<->|] (-1,-1.25) -- (0.95,-1.25) node[midway,below] {\scalebox{0.9}{$t-\ell$}};
    \draw[|<->|] (1.05,-1.25) -- (3,-1.25) node[midway,below] {\scalebox{0.9}{$2\ell-t$}};
  \end{tikzpicture}
  = 2^{(1+n)\ell-2nt}
  \begin{tikzpicture}[baseline={([yshift=-0.6ex]current bounding box.center)},scale=0.5]
    \prop{0}{1}{FncolU}
    \prop{1}{2}{FncolU}
    \prop{2}{3}{FncolU}
    \prop{3}{4}{FncolU}
    \prop{1}{0}{FncolU}
    \prop{2}{1}{FncolU}
    \prop{3}{2}{FncolU}
    \fSME{-0.5}{1.5}
    \fSME{0.5}{2.5}
    \fSME{1.5}{3.5}
    \fSME{2.5}{4.5}
    \fME{3.5}{4.5}
    \fME{3.5}{3.5}
    \fME{3.5}{2.5}
    \fME{3.5}{1.5}
    \MEld{2.5}{0.5}
    \MEld{1.5}{-0.5}
    \fME{0.5}{-0.5}
    \fME{-0.5}{0.5}
  \end{tikzpicture},
\end{equation}
where the last equality follows from the realization that $\PdFlattext$ is up to normalization
equal to $\fMEtext$.
To simplify the rest, we need to introduce a projector to the subspace of products of 
two-site states $[1,0,0,1]$ and $[0,1,1,0]$ in the copies connected by $\fMEtext$
\begin{equation}
  \begin{tikzpicture}[baseline={([yshift=-0.6ex]current bounding box.center)},scale=0.5]
    \gridLine{-0.3}{-0.3}{0.3}{0.3}
    \PQ{0}{0}
  \end{tikzpicture}=
  \frac{1}{2}
  \begin{bmatrix}
    1&0&0&1\\
    0&1&1&0\\
    0&1&1&0\\
    1&0&0&1
  \end{bmatrix}^{\otimes n},
\end{equation}
which can be shown to obey the following
\begin{equation} \label{eq:relationsProjectorQ}
  \begin{tikzpicture}[baseline={([yshift=-0.6ex]current bounding box.center)},scale=0.5]
    \draw[white] (0,0.75) -- (0,-0.75);
    \gridLine{0}{0}{-0.75}{-0.75}
    \prop{0}{0}{FncolU}
    \PQ{-0.5}{-0.5}
  \end{tikzpicture}=
  \begin{tikzpicture}[baseline={([yshift=-0.6ex]current bounding box.center)},scale=0.5]
    \draw[white] (0,0.75) -- (0,-0.75);
    \gridLine{0}{0}{-0.75}{0.75}
    \prop{0}{0}{FncolU}
    \PQ{-0.5}{0.5}
  \end{tikzpicture},\qquad
  \begin{tikzpicture}[baseline={([yshift=-0.6ex]current bounding box.center)},scale=0.5]
    \draw[white] (0,0.75) -- (0,-0.75);
    \gridLine{0}{0}{-0.75}{-0.75}
    \gridLine{0}{0}{0.75}{-0.75}
    \prop{0}{0}{FncolU}
    \PQ{-0.5}{-0.5}
    \PQ{0.5}{-0.5}
  \end{tikzpicture}=
  \begin{tikzpicture}[baseline={([yshift=-0.6ex]current bounding box.center)},scale=0.5]
    \draw[white] (0,0.75) -- (0,-0.75);
    \gridLine{0}{0}{-0.75}{0.75}
    \gridLine{0}{0}{0.75}{0.75}
    \prop{0}{0}{FncolU}
    \PQ{-0.5}{0.5}
    \PQ{0.5}{0.5}
  \end{tikzpicture},\qquad
  \begin{tikzpicture}[baseline={([yshift=-0.6ex]current bounding box.center)},scale=0.5]
    \gridLine{-0.3}{-0.3}{0.3}{0.3}
    \fME{-0.3}{-0.3}
  \end{tikzpicture}=
  \begin{tikzpicture}[baseline={([yshift=-0.6ex]current bounding box.center)},scale=0.5]
    \gridLine{-0.3}{-0.3}{0.3}{0.3}
    \PQ{0}{0}
    \fME{-0.3}{-0.3}
  \end{tikzpicture},\qquad
  \begin{tikzpicture}[baseline={([yshift=-0.6ex]current bounding box.center)},scale=0.5]
    \gridLine{-0.3}{-0.3}{0.3}{0.3}
    \MErd{-0.3}{-0.3}
  \end{tikzpicture}=
  \begin{tikzpicture}[baseline={([yshift=-0.6ex]current bounding box.center)},scale=0.5]
    \gridLine{-0.3}{-0.3}{0.3}{0.3}
    \PQ{0}{0}
    \MErd{-0.3}{-0.3}
  \end{tikzpicture}.
\end{equation}
Using it we can formulate a $n$-replica generalization of the hierarchy-$2$ relation
\begin{equation}
  \begin{tikzpicture}[baseline={([yshift=-0.6ex]current bounding box.center)},scale=0.5]
    \gridLine{0}{0}{0.75}{-0.75}
    \prop{0}{0}{FncolU}
    \prop{1}{1}{FncolU}
    \PQ{0.5}{-0.5}
    \fME{-0.5}{-0.5}
    \fSME{-0.5}{0.5}
    \fSME{0.5}{1.5}
  \end{tikzpicture}=
  \begin{tikzpicture}[baseline={([yshift=-0.6ex]current bounding box.center)},scale=0.5]
    \gridLine{0.125}{-0.125}{0.75}{-0.75}
    \prop{1}{1}{FncolU}
    \PQ{0.5}{-0.5}
    \fSME{0.5}{1.5}
    \fME{0.5}{0.5}
    \fSME{0.125}{-0.125}
  \end{tikzpicture}.
  \label{eq:gen2h}
\end{equation}
which is proven in Sec.~\ref{sec:proofgen2h}. Repeatedly applying it together with~\eqref{eq:relationsProjectorQ}, it simplifies the expression~\eqref{eq:FlatStateIntermediate2} into
\begin{equation}
  \tr[\rho_{A}^n(t)]
  = 2^{(1+n)\ell-2nt}
  \left(\begin{tikzpicture}[baseline={([yshift=-0.6ex]current bounding box.center)},scale=0.5]
    \gridLine{0}{0}{0}{0.75}
    \MEh{0}{0}
    \fSME{0}{0.75}
  \end{tikzpicture}\right)^{2\ell-t}
  \left(\begin{tikzpicture}[baseline={([yshift=-0.6ex]current bounding box.center)},scale=0.5]
    \gridLine{0}{0}{0}{0.75}
    \fME{0}{0}
    \fSME{0}{0.75}
  \end{tikzpicture}\right)^{t-\ell}
  \left(\begin{tikzpicture}[baseline={([yshift=-0.6ex]current bounding box.center)},scale=0.5]
    \gridLine{0}{0}{0}{0.75}
    \fME{0}{0}
    \fME{0}{0.75}
  \end{tikzpicture}\right)^{t-\ell}
  =2^{(1-n)t}.
\end{equation}

\subsubsection{Proof of Eq.~(\ref{eq:gen2h})}
\label{sec:proofgen2h}
To prove \eqref{eq:gen2h} we expand the leg connecting the two gates on the l.h.s.\ in the basis 
\begin{equation}
  \ket{\alpha_1,\ldots,\alpha_n}=\begin{tikzpicture}[baseline={([yshift=0ex]current bounding box.center)},scale=0.5]
    \gridLine{0}{0}{0}{0.75}
    \fME{0}{0}
    \node at (0.25,-0.35) {\scalebox{0.7}{$\alpha_1$}};
  \end{tikzpicture}\!\!\!\!\!
  \otimes \cdots \otimes
  \!\!\begin{tikzpicture}[baseline={([yshift=0ex]current bounding box.center)},scale=0.5]
    \gridLine{0}{0}{0}{0.75}
    \fME{0}{0}
    \node at (0.25,-0.35) {\scalebox{0.7}{$\alpha_n$}};
  \end{tikzpicture} \equiv \begin{tikzpicture}[baseline={([yshift=0ex]current bounding box.center)},scale=0.5]
    \gridLine{0}{0}{0}{0.75}
    \fME{0}{0}
    \node at (0.25,-0.35) {\scalebox{0.7}{$\alpha_1,\ldots, \alpha_n$}};
  \end{tikzpicture}, \qquad \alpha_j = 0,1,2,3, 
\end{equation}
where 
\begin{equation}
  \begin{tikzpicture}[baseline={([yshift=0ex]current bounding box.center)},scale=0.5]
    \gridLine{0}{0}{0}{0.75}
    \fME{0}{0}
    \node at (0.25,-0.35) {\scalebox{0.7}{$\alpha$}};
  \end{tikzpicture}
  =
  \begin{tikzpicture}[baseline={([yshift=0ex]current bounding box.center)},scale=0.5]
    \tbotHook{0.3}{0}{0.15}
    \tgridLine{0}{0}{0}{1}
    \tgridLine{0.3}{0}{0.3}{1}
    \Pd{0.15}{-0.1}
    \node at (0.5,-0.5) {\scalebox{0.7}{$\sigma^{(\alpha)}$}};
  \end{tikzpicture},
\end{equation}
and $\sigma^{(\alpha)}$ represents the $\alpha$-th Pauli matrix ($\sigma^{(0)}=I$). Observing that 
\begin{equation} \label{eq:simprel}
  \begin{tikzpicture}[baseline={([yshift=-0.6ex]current bounding box.center)},scale=0.5]
    \gridLine{0}{0}{0.75}{-0.75}
    \prop{0}{0}{FcolU}
    \fME{-0.5}{-0.5}
    \PQ{0.5}{-0.5}
  \end{tikzpicture} =
  \begin{tikzpicture}[baseline={([yshift=-1.2ex]current bounding box.center)},scale=0.5]
    \gridLine{-0.25}{0.25}{-0.25}{0.75}
    \fME{-0.25}{0.25}
  \end{tikzpicture}\otimes \begin{tikzpicture}[baseline={([yshift=-0.6ex]current bounding box.center)},scale=0.5]
    \gridLine{-0.25}{-0.25}{-0.25}{-0.75}
    \gridLine{-0.25}{0.25}{-0.25}{0.75}
    \fME{-0.25}{-0.25}
    \fME{-0.25}{0.25}
  \end{tikzpicture}
  + \begin{tikzpicture}[baseline={([yshift=-0.2ex]current bounding box.center)},scale=0.5]
    \gridLine{-0.25}{0.25}{-0.25}{0.75}
    \fME{-0.25}{0.25}
    \node at (-0.125,-0.1) {\scalebox{0.7}{$1$}};
  \end{tikzpicture}\otimes \begin{tikzpicture}[baseline={([yshift=-0.6ex]current bounding box.center)},scale=0.5]
    \gridLine{-0.25}{-0.25}{-0.25}{-0.75}
    \gridLine{-0.25}{0.25}{-0.25}{0.75}
    \fME{-0.25}{-0.25}
    \fME{-0.25}{0.25}
    \node at (0.1, 0.2) {\scalebox{0.7}{$1$}};
    \node at (0.1,-0.2) {\scalebox{0.7}{$1$}};
  \end{tikzpicture}, \qquad
  \begin{tikzpicture}[baseline={([yshift=-0.6ex]current bounding box.center)},scale=0.5]
    \gridLine{0}{0}{0.75}{-0.75}
    \prop{0}{0}{FcolU}
    \fME{-0.5}{-0.5}
    \node at (-0.375,-0.85) {\scalebox{0.7}{$1$}};
    \PQ{0.5}{-0.5}
  \end{tikzpicture} =
  \begin{tikzpicture}[baseline={([yshift=-1.2ex]current bounding box.center)},scale=0.5]
    \gridLine{-0.25}{0.25}{-0.25}{0.75}
    \fME{-0.25}{0.25}
  \end{tikzpicture}\otimes \begin{tikzpicture}[baseline={([yshift=-0.6ex]current bounding box.center)},scale=0.5]
    \gridLine{-0.25}{-0.25}{-0.25}{-0.75}
    \gridLine{-0.25}{0.25}{-0.25}{0.75}
    \fME{-0.25}{-0.25}
    \fME{-0.25}{0.25}
    \node at (0.1, 0.2) {\scalebox{0.7}{$1$}};
    \node at (0.1,-0.2) {\scalebox{0.7}{$1$}};
  \end{tikzpicture}+
  \begin{tikzpicture}[baseline={([yshift=-0.2ex]current bounding box.center)},scale=0.5]
    \gridLine{-0.25}{0.25}{-0.25}{0.75}
    \fME{-0.25}{0.25}
    \node at (-0.125,-0.1) {\scalebox{0.7}{$1$}};
  \end{tikzpicture}\otimes \begin{tikzpicture}[baseline={([yshift=-0.6ex]current bounding box.center)},scale=0.5]
    \gridLine{-0.25}{-0.25}{-0.25}{-0.75}
    \gridLine{-0.25}{0.25}{-0.25}{0.75}
    \fME{-0.25}{-0.25}
    \fME{-0.25}{0.25}
  \end{tikzpicture},
\end{equation}
we have that the l.h.s.\ of Eq.~\eqref{eq:gen2h} is written as 
\begin{equation} \label{eq:intermediatestep1}
  \begin{tikzpicture}[baseline={([yshift=-0.6ex]current bounding box.center)},scale=0.5]
    \gridLine{0}{0}{0.75}{-0.75}
    \prop{0}{0}{FncolU}
    \prop{1}{1}{FncolU}
    \PQ{0.5}{-0.5}
    \fME{-0.5}{-0.5}
    \fSME{-0.5}{0.5}
    \fSME{0.5}{1.5}
  \end{tikzpicture}= \frac{1}{2^n}
  \sum_{\alpha_j=0,1,2,3}
  \begin{tikzpicture}[baseline={([yshift=-0.6ex]current bounding box.center)},scale=0.5]
    \gridLine{0}{0}{0.75}{-0.75}
    \prop{0}{0}{FncolU}
    \prop{1.5}{1.5}{FncolU}
    \PQ{0.5}{-0.5}
    \fME{-0.5}{-0.5}
    \fME{0.5}{0.5}
    \fME{1}{1}
    \fSME{-0.5}{0.5}
    \fSME{1}{2}
    \node at (1.75,0.25) {\scalebox{0.7}{$\alpha_1,\ldots, \alpha_n$}};
    \node at (2,0.6) {\scalebox{0.7}{$\alpha_1,\ldots, \alpha_n$}};
  \end{tikzpicture}= \frac{1}{2^n}
  \sum_{\alpha_j=0,1}
  \begin{tikzpicture}[baseline={([yshift=-0.6ex]current bounding box.center)},scale=0.5]
    \gridLine{0}{0}{0.75}{-0.75}
    \prop{0}{0}{FncolU}
    \prop{1.5}{1.5}{FncolU}
    \PQ{0.5}{-0.5}
    \fME{-0.5}{-0.5}
    \fME{0.5}{0.5}
    \fME{1}{1}
    \fSME{-0.5}{0.5}
    \fSME{1}{2}
    \node at (1.75,0.25) {\scalebox{0.7}{$\alpha_1,\ldots, \alpha_n$}};
    \node at (2,0.6) {\scalebox{0.7}{$\alpha_1,\ldots, \alpha_n$}};
  \end{tikzpicture}.
\end{equation}
Next we note that Eq.~\eqref{eq:simprel} implies
\begin{equation}
  \begin{tikzpicture}[baseline={([yshift=-0.6ex]current bounding box.center)},scale=0.5]
    \gridLine{0}{0}{0.75}{-0.75}
    \prop{0}{0}{FcolU}
    \fME{-0.5}{-0.5}
    \node at (-0.375,-0.85) {\scalebox{0.7}{$\alpha$}};
    \node at (-0.375,0.85) {\scalebox{0.7}{\phantom{$\alpha$}}};
    \PQ{0.5}{-0.5}
  \end{tikzpicture}=
  \mkern-28mu
  \begin{tikzpicture}[baseline={([yshift=-0.6ex]current bounding box.center)},scale=0.5]
    \tgridLine{-0.6}{0.7}{-0.6}{1.1}
    \tgridLine{-0.4}{0.7}{-0.4}{1.1}
    \gridLine{0.5}{0.7}{0.5}{1.1}
    \Pd{-0.6}{0.9}
    \node at (-1.2, 1.3) {\scalebox{0.7}{$\sigma^{(\alpha)}$}};
    \node at (-1.2, -1.3) {\scalebox{0.7}{\phantom{$\sigma^{(\alpha)}$}}};
    \gridLine{0}{0}{0.75}{-0.75}
    \prop{0}{0}{FcolU}
    \fME{-0.5}{-0.5}
    \PQ{0.5}{-0.5}
  \end{tikzpicture}=
  \begin{tikzpicture}[baseline={([yshift=-0.6ex]current bounding box.center)},scale=0.5]
    \tgridLine{-0.6}{0.7}{-0.6}{1.1}
    \tgridLine{-0.4}{0.7}{-0.4}{1.1}
    \gridLine{0.5}{0.7}{0.5}{1.1}
    \Pd{-0.4}{0.9}
    \node at (0.2, 1.3) {\scalebox{0.7}{$\sigma^{(\alpha)}$}};
    \node at (0.2, -1.3) {\scalebox{0.7}{\phantom{$\sigma^{(\alpha)}$}}};
    \gridLine{0}{0}{0.75}{-0.75}
    \prop{0}{0}{FcolU}
    \fME{-0.5}{-0.5}
    \PQ{0.5}{-0.5}
  \end{tikzpicture}, \qquad \alpha=0,1\,.
\end{equation}
This means that 
\begin{equation}
  \forall \alpha_j =0,1,\qquad\qquad 
  \begin{tikzpicture}[baseline={([yshift=-0.6ex]current bounding box.center)},scale=0.5]
    \prop{1.5}{1.5}{FncolU}
    \fME{1}{1}
    \fSME{1}{2}
    \node at (2,0.6) {\scalebox{0.7}{$\alpha_1,\ldots, \alpha_n$}};
  \end{tikzpicture} = \begin{cases} 
    \begin{tikzpicture}[baseline={([yshift=-0.6ex]current bounding box.center)},scale=0.5]
      \prop{0}{0}{FncolU}
      \fME{-0.5}{-0.5}
      \fSME{-0.5}{0.5}
    \end{tikzpicture} & \sum_j \alpha_j \,\, {\rm even} \\
    \\
    \begin{tikzpicture}[baseline={([yshift=-0.6ex]current bounding box.center)},scale=0.5]
      \prop{0}{0}{FncolU}
      \fME{-0.5}{-0.5}
      \fSME{-0.5}{0.5} 
      \node at (0.5,-1) {\scalebox{0.7}{$1,0\ldots, 0$}};
    \end{tikzpicture}
    & \sum_j \alpha_j \,\,{\rm odd} \\
  \end{cases}\,.
\end{equation}
Plugging back into Eq.~\eqref{eq:intermediatestep1} we then find 
\begin{equation}
  \begin{tikzpicture}[baseline={([yshift=-0.6ex]current bounding box.center)},scale=0.5]
    \gridLine{0}{0}{0.75}{-0.75}
    \prop{0}{0}{FncolU}
    \prop{1}{1}{FncolU}
    \PQ{0.5}{-0.5}
    \fME{-0.5}{-0.5}
    \fSME{-0.5}{0.5}
    \fSME{0.5}{1.5}
  \end{tikzpicture}= 
    \sum_{\substack{\alpha_j=0,1 \\ \sum_j \alpha_j {\rm even}}}
    \begin{tikzpicture}[baseline={([yshift=-0.6ex]current bounding box.center)},scale=0.5]
      \gridLine{-0.5}{0.5}{-0.5}{-0.25}
      \gridLine{-0.5}{-1.25}{-0.5}{-1.75}
      \prop{1.5}{-0.5}{FncolU}
      \fME{1}{-1}
      \fSME{1}{0}
      \fSME{-0.5}{0.5}
      \fME{-0.5}{-0.25}
      \fME{-0.5}{-1.25}
      \node at (-.5,-0.5) {\scalebox{0.7}{$\alpha_1,\ldots, \alpha_n$}};
      \node at (-.5,-0.95) {\scalebox{0.7}{$\alpha_1,\ldots, \alpha_n$}};
    \end{tikzpicture}+
    \sum_{\substack{\alpha_j=0,1 \\ \sum_j \alpha_j {\rm odd}}}
   \begin{tikzpicture}[baseline={([yshift=-0.6ex]current bounding box.center)},scale=0.5]
    \gridLine{-0.5}{0.5}{-0.5}{-0.25}
    \gridLine{-0.5}{-1.25}{-0.5}{-1.75}
    \prop{1.5}{-0.5}{FncolU}
    \fME{1}{-1}
    \fSME{1}{0}
    \fSME{-0.5}{0.5}
    \fME{-0.5}{-0.25}
     \fME{-0.5}{-1.25}
     \node at (-.5,-0.5) {\scalebox{0.7}{$\alpha_1,\ldots, \alpha_n$}};
     \node at (-.5,-0.95) {\scalebox{0.7}{$\alpha_1,\ldots, \alpha_n$}};
      \node at (2,-1.5) {\scalebox{0.7}{$1,0\ldots, 0$}};
  \end{tikzpicture},
\end{equation}
where we used Eq.~\eqref{eq:simprel}. Finally, noting 
\begin{equation}
 \sum_{\substack{\alpha_j=0,1 \\ \sum_j \alpha_j {\rm even}}}
   \begin{tikzpicture}[baseline={([yshift=-0.6ex]current bounding box.center)},scale=0.5]
    \gridLine{-0.5}{0.5}{-0.5}{-0.25}
    \gridLine{-0.5}{-1.25}{-0.5}{-1.75}
    \fSME{-0.5}{0.5}
    \fME{-0.5}{-0.25}
     \fME{-0.5}{-1.25}
     \node at (-.5,-0.5) {\scalebox{0.7}{$\alpha_1,\ldots, \alpha_n$}};
     \node at (-.5,-0.95) {\scalebox{0.7}{$\alpha_1,\ldots, \alpha_n$}};
  \end{tikzpicture} =  \sum_{\alpha_j=0,1}
   \begin{tikzpicture}[baseline={([yshift=-0.6ex]current bounding box.center)},scale=0.5]
    \gridLine{-0.5}{0.5}{-0.5}{-0.25}
    \gridLine{-0.5}{-1.25}{-0.5}{-1.75}
    \fSME{-0.5}{0.5}
    \fME{-0.5}{-0.25}
     \fME{-0.5}{-1.25}
     \node at (-.5,-0.5) {\scalebox{0.7}{$\alpha_1,\ldots, \alpha_n$}};
     \node at (-.5,-0.95) {\scalebox{0.7}{$\alpha_1,\ldots, \alpha_n$}};
  \end{tikzpicture} =     \begin{tikzpicture}[baseline={([yshift=-0.6ex]current bounding box.center)},scale=0.5]
    \gridLine{-0.5}{0.5}{0.5}{-0.5}
    \fSME{-0.5}{0.5}
    \PQ{0}{0}
  \end{tikzpicture},
  \qquad
\sum_{\substack{\alpha_j=0,1 \\ \sum_j \alpha_j {\rm odd}}}
   \begin{tikzpicture}[baseline={([yshift=-0.6ex]current bounding box.center)},scale=0.5]
    \gridLine{-0.5}{0.5}{-0.5}{-0.25}
    \gridLine{-0.5}{-1.25}{-0.5}{-1.75}
    \fSME{-0.5}{0.5}
    \fME{-0.5}{-0.25}
     \fME{-0.5}{-1.25}
     \node at (-.5,-0.5) {\scalebox{0.7}{$\alpha_1,\ldots, \alpha_n$}};
     \node at (-.5,-0.95) {\scalebox{0.7}{$\alpha_1,\ldots, \alpha_n$}};
  \end{tikzpicture} = 0, 
\end{equation}
we obtain the r.h.s.\ of Eq.~\eqref{eq:gen2h}. 

\end{document}